# Berry curvature induced giant anomalous and spin texture driven Hall responses in the layered kagome antiferromagnet GdTi₃Bi₄

Shobha Singh,[1] Shivam Rathod,[1] Rong chen[2], Lipika[1], Sneh[1], Rie Y. Umetsu[3,4], Yan Sun[2] and Kaustuv Manna[1]*

[1]*Department of Physics, Indian Institute of Technology Delhi, New Delhi 110016, India*
[2]*Shenyang National Laboratory for Materials Science, Institute of Metal Research, Chinese Academy of Sciences 72 Wenhua Road, Shenyang 110016, China.*
[3]*Institute for Materials Research, Tohoku University, Sendai 980-8577, Japan*
[4]*Center for Science and Innovation in Spintronics, Tohoku University, Sendai 980-8577, Japan*

* E-mail: kaustuvmanna@iitd.ac.in

In recent years, layered kagome magnets have emerged as promising platforms for Berry-curvature engineering and unconventional transport phenomena. Here, we present the single-crystal growth, magnetization, and electrical transport characterizations of the van der Waals-like layered antiferromagnet GdTi₃Bi₄. The system exhibits pronounced field-induced first-order phase transitions. Comprehensive frequency, temperature, and field-dependent ac susceptibility measurements, and Hall analysis, reveals the formation of a spin-cluster-like glassy magnetic phase attributed to noncollinear spin textures. Additionally, the system demonstrates a colossal anomalous Hall conductivity ($\sigma_{xy}^A \approx 8.6(7) \times 10^3 \ \Omega^{-1} \ cm^{-1}$ at 2 K). Detailed scaling analyses reveal the coexistence of skew scattering and intrinsic Berry-curvature contributions to the anomalous Hall effect. First-principles calculations highlight flat-band near the Fermi level, with $f$-electrons of the Gd ion contributing large intrinsic Hall response. Thus, GdTi₃Bi₄ emerges as a rare layered kagome magnet, exhibiting Berry curvature-induced giant anomalous and spin texture-driven Hall responses, providing a versatile platform for exploring spin-texture physics and advancing low-dimensional spintronic functionalities.

*Introduction*.

Topological quantum materials, which exhibit significant Berry curvature in both real and momentum spaces, represent a prominent area of contemporary research. These materials are distinguished by their rare dual Hall response, integrating both anomalous and topological Hall effects within a cohesive framework. In this context, the intrinsic anomalous Hall conductivity (AHC) is predominantly influenced by the Berry curvature, a geometric characteristic of the electronic bands. In materials with strong spin-orbit coupling or nontrivial band topology, particularly those with band crossings or entangled bands near the Fermi energy, a substantial Berry curvature-induced AHC is observed [1–3]. Similarly, a large real-space Berry curvature can lead to pronounced topological Hall effects, often associated with skyrmion lattices or non-trivial spin textures, which generate effective magnetic fields for conduction electrons [4–9]. However, materials exhibiting robust dual Hall responses remain rare.

Among the promising materials, the recently identified two-dimensional kagome lattice compound family, $R$Ti$_3$Bi$_4$ ($R$: rare earth) is particularly noteworthy. These materials are characterized by layers that are weakly bonded through van der Waals interactions, which facilitates their exfoliation into atomically thin sheets [10] and potential technological applications in spintronics, spin-orbit torque devices, and Hall-based field sensors, among others [11,12]. GdTi$_3$Bi$_4$ is an example of such a layered antiferromagnet with an orthorhombic structure, featuring Gd zigzag chains and Ti-based kagome layers. This distinctive structural feature results in pronounced magnetic anisotropy and multiple field-induced magnetic phases.

The previous studies on GdTi$_3$Bi$_4$ have reported the formation of 1/3 magnetization plateaus, complex phase diagrams and real-space imaging of spin-texture domains [13–18]. In the present work, we performed a comprehensive analysis of dc magnetization, ac susceptibility, and temperature-field dependent electrical transport measurements on GdTi$_3$Bi$_4$ single crystals. Our findings reveal that GdTi$_3$Bi$_4$ is a highly anisotropic, frustrated antiferromagnet, characterized by various field-induced magnetic states, slow spin dynamics typical of magnetic glass at the second metamagnetic transition ($H_{C2}$ = 3.4 T), and notable magnetoresistance jumps akin to those observed in giant magnetoresistance (GMR) multilayers. We performed systematic decomposition of the Hall resistivity data into ordinary, anomalous, and additional spin-texture-related terms at low temperature, which comprises both extrinsic and Berry curvature-induced intrinsic contributions. At low temperatures, we identified dual Hall responses: a giant anomalous Hall conductivity and a significant spin texture-induced Hall resistivity. We employed the Tian–Ye–Jin (TYJ) model to extract the extrinsic and intrinsic anomalous Hall contributions. Theoretical calculations suggest that the Hall signal is further amplified by the $f$-electron magnetic effects of the Gd atoms. Collectively, these findings identify GdTi$_3$Bi$_4$ as a unique and versatile candidate demonstrating the coexistence of large anomalous and significant spin texture driven Hall conductivities originating from the momentum space and the real space Berry curvature effect in a layered kagome lattice antiferromagnet.

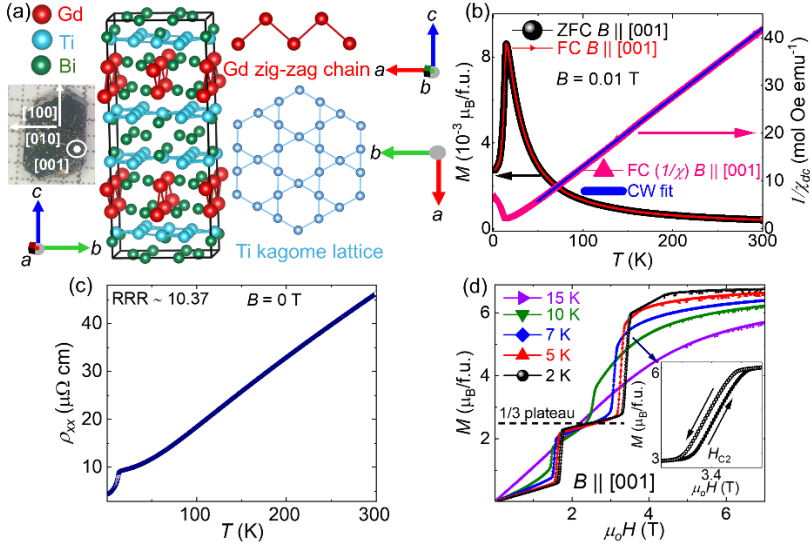

**Figure 1.** (a) Crystal structure of GdTi$_3$Bi$_4$, with the Gd zig-zag chain and Ti kagome layer, optical image of as-grown single crystal with crystallographic orientations is shown in the left. (b) ZFC and FC dc magnetization $M(T)$ at 0.01 T magnetic field on left axis and dc $1/\chi(T)$ on right axis measured at 0.01 T magnetic field along c-axis. The blue solid line represents the CW fit. (c) Temperature dependent resistivity, revealing overall metallic behaviour and additional hump around the AFM transition. d) Isothermal magnetization $M(H)$ for $B \parallel c$ at different temperatures, the inset shows zoomed in view of the hysteresis at $H_{C2} \sim 3.4$ T at 2 K.

GdTi$_3$Bi$_4$ single crystals were synthesized using flux method, as detailed in the Supporting-Information (SI) [19]. The high crystallinity of the grown crystals was confirmed through a comprehensive analysis involving XRD, SEM, EDX, and Laue techniques (Figures S1-S3 in SI [19]). The determined crystal structure, along with an image of one of the synthesized crystals, is presented in Figure 1(a). Figure 1(b) illustrates the temperature-dependent zero-field-cooled (ZFC) and field-cooled (FC) dc magnetization ($M(T)$) measured with the magnetic field applied along [001] crystallographic direction ($M(T)$ along [100] and [010] are shown in SI Figure S4(a) [19]). $M(T)$ exhibits a peak at $T_N$ = 14.2 K, signifying the onset of antiferromagnetic ordering within the compound. We observed an additional kink at $T_{N2}$ = 13.2 K (highlighted in SI Figure S4(b) [19]), likely due to the spin reorientation process, as reported in other Gd-based systems [13,32] as shown. The ZFC and FC curves completely overlap down to the lowest measured temperature indicating the absence of magnetic irreversibility. The field-cooled dc inverse susceptibility $1/\chi(T)$ is depicted in Figure 1b for $B \parallel$ [001]. In the paramagnetic regime, $1/\chi_{dc}(T)$ follows behaviour described by the Weiss molecular field theory for local-moment antiferromagnets [33]. Fitting the high-temperature linear region ($T > 50$ K) using the Curie-Weiss (CW) law, $\chi(T) = C/(T - \theta_p)$, yields an effective magnetic moment, $\mu_{eff}$ = 7.4 $\mu_B$ (slope) and a Curie-Weiss temperature $\theta_p$ = 9.81 K (intercept). The corresponding CW fit along [100] and [010] is shown in Figures S5(a, b) respectively in SI [19]. The positive $\theta_p$ indicates FM interaction between the Gd ions at higher temperatures, consistent with previous reports [14,17]. Magnetization saturates at a relatively lower field for $B \parallel$ [001] compared to the other directions, indicating that the $c$ is the easy axis of magnetization as shown in SI Figure S6(a, b) [19]. The temperature-dependent electrical resistivity, $\rho_{xx}(T)$, (as depicted in Figure 1(c) reveals an overall metallic behaviour in GdTi$_3$Bi$_4$ crystals, characterized by a nearly linear decrease in resistivity with decreasing temperature. Notably, a hump-like feature is observed around $T_N$, followed by a sharp decline in $\rho_{xx}$ upon further cooling. This behaviour is likely attributable to the suppression of spin-disorder scattering in the compound below $T_N$ [34]. The residual resistivity ratio (RRR) is determined to be $\rho_{xx}$ (300 K)/$\rho_{xx}$ (2 K)

= 10.37. This comparatively higher RRR value [13,15], coupled with a low residual resistivity $\rho_{xx}$ (2 K) of 4.46 $\mu\Omega$ cm, indicates good crystalline quality of the synthesized single crystals. The first principles-based band structure calculations are shown in SI Figure S7 (a-c) [19].

Magnetisation dynamics in GdTi$_3$Bi$_4$ were further examined through comprehensive dc and ac magnetisation experiments, with the magnetic field applied along the easy axis, $B \parallel [001]$. The magnetization as a function of the magnetic field $M(H)$, measured along the easy axis [001], is depicted in Figure 1(d). At 2 K, $M(H)$ initially increases linearly with the applied field, demonstrating characteristic AFM behaviour. At 1.7 T ($H_{C1}$), a pronounced increase in magnetization occurs, signifying a field-induced metamagnetic transition (MMT). Subsequently, a 1/3 magnetization plateau is observed, followed by a second MMT at 3.4 T ($H_{C2}$). Beyond $H_{C2}$, a third transition around 4.5 T indicates a fully field-polarized state (for measurements along [100] and [010], refer Figure S8 and S9 in SI [19]). Similar MMTs have been observed in other Gd-based compounds [6,32,35]. The emergence of MMTs may be attributed to multiple magnetic lattices with non-collinear spins, or to geometrical frustrations in triangular lattice AFMs [6,36]. In AFM layers with strong anisotropy, spins do not directly transition from out-of-plane to in-plane or vice versa; rather, they undergo a gradual process of spin polarization, adopting various spin configurations. During this process, a specific spin arrangement compatible with the lattice geometry can be stabilized within certain magnetic field regimes. In a kagome AFM with spin-frustrated triangles, this results in the formation of an up-up-down spin phase, manifested as a magnetization plateau at one-third of its saturation magnetization [37,38].

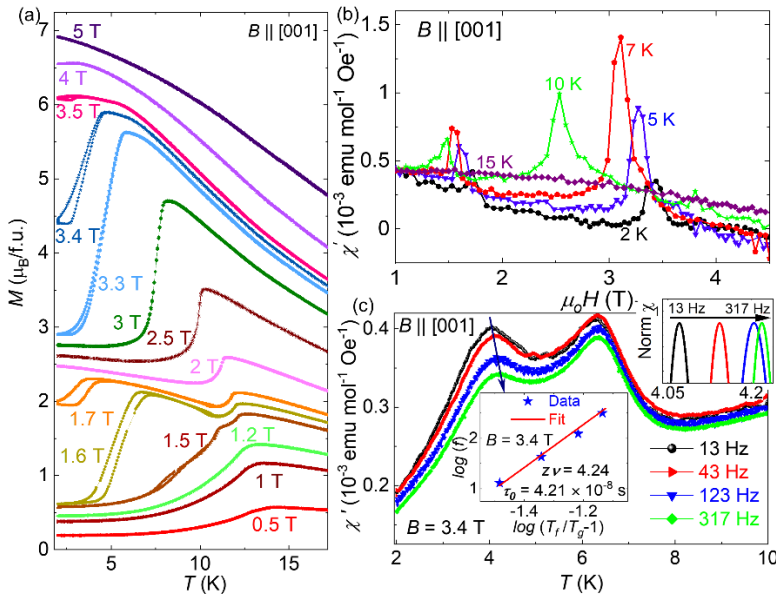

**Figure 2.** (a) Temperature dependent FCC and FCW magnetization curves at various magnetic fields for $B \parallel c$. (b) Field-dependent ac susceptibility $\chi'(H)$ measured at different temperatures. (c) In-phase component of temperature-dependent ac susceptibility $\chi'(T)$ measured at various frequencies shown for $H_{ac} = 9$ Oe and $H_{dc} = 3.4$ T. The top inset provides a corresponding zoomed-in view of the normalized $\chi'(T)$ at various frequencies as listed in panel. Initially, the experimental data were fitted with a polynomial, then normalized and magnified for enhanced visualization of the frequency-dependent peak shift. The bottom inset presents the power-law fitting demonstrating the variation of $T_f$ with change in applied frequency.

The two MMTs occurring at critical fields $H_{C1} \sim 1.7$ T and $H_{C2} \sim 3.4$ T exhibit narrow hysteresis loops during field increase and field decrease (see inset of Figure 1d), indicating a field induced first-order magnetic phase transition [39]. As the temperature rises, these hysteresis loops shift towards lower fields due to a reduction in stiffness of magnetic interactions and eventually disappearing above $T_N$. To further substantiate the first-order nature of these transitions, temperature dependent magnetization was recorded under various magnetic fields, as depicted in Figure 2(a). Around the $H_{C1} \sim 1.7$ T and $H_{C2} \sim 3.4$ T, thermal hysteresis is observed between the field-cooled cooling (FCC) and field-cooled warming (FCW) magnetization curves, indicating a first order phase transition and competing magnetic orders in the system [40]. The irreversibility around $H_{C1} \sim 1.7$ T and $H_{C2} \sim 3.4$ T, followed by the overlap of FCC and FCW upon further temperature decrease, is reminiscent of the kinetic arrest of a first order phase transition in the system. Such non-equilibrium state generally results in frustrated magnetism, akin the magnetic glass [41,42].

To elucidate the origin of the first order phase transition observed in the $M(H)$ measurements, we conducted field-dependent ac susceptibility ($\chi'(H)$) measurements. The real (in-phase) component $\chi'$ is depicted in Figure 2(b). A variable dc field was superimposed with a constant ac magnetic field of amplitude 9 Oe and an excitation frequency 13 Hz. At 2 K, the $\chi'(H)$ data reveal two peaks and a minor dip. The peak in the $\chi'(H)$ measurement shifts towards a lower magnetic field as the temperature increases, eventually disappearing at $T_N$, in agreement with the dc $M(H)$ measurements. The magnetic fields at which these field-induced transitions occur are consistent across both dc $M(H)$ and $\chi'(H)$ measurements. To probe the associated spin dynamics, temperature-dependent ac susceptibility $\chi'(T)$ was measured at various excitation frequencies of ac amplitude 9 Oe. Figure 2(c) presents the $\chi'(T)$ measured with a superimposed dc field $H_{dc} = 3.4$ T, corresponding to the second MMT. Two distinct peaks are observed near 3.5 and 6 K, with increasing dc field, $T_N$ shifts to a lower temperature. Interestingly, the peak around 6 K remains invariant with varying excitation frequency ($f$), while the peak around 3.5 K shifts toward higher temperatures as $f$ increases (Figure 2(c) inset). This behaviour is the characteristic of magnetic frustration or a glassy magnetic state, where $T_f$ linearly shifts with log ($f$) (validated in SI Figure S10 [19]) [43]. In contrast, $\chi'(T)$ with $H_{dc} = 0$, 1.7 T (across the first MMT), and 3.95 T do not exhibit any frequency dependent shift, indicating the absence of any glassy transition or magnetic frustrations (see SI Figure S10 [19]). The frequency-dependent peak shift at 3.4 T, was scaled using the critical slowing down model, [44] expressed as $\tau = \tau_o \left( T_f / T_g - 1 \right)^{-zv}$ where $\tau = 1/f$, $\tau_o$ is the spin relaxation time, $T_f$ is the spin freezing temperature corresponding to the excitation frequency $f$, $zv$ is the dynamical critical constant, and $T_g$ is the spin freezing temperature for $f \to 0$. Initially, $T_g$ was estimated from the linear fit of $T_f$ vs $log$ ($f$) in the limit $f \to 0$ as 3.96 $K$ (see SI Figure S10 [19]). Finally, $\tau_o$ and $zv$ were estimated from the intercept and slope, respectively, using the linear fit of the curve $log_{10}(f)$ vs $log_{10}(T_f / T_g - 1)$ as shown in the inset of Figure 2c, yielding $\tau_o \approx 4.21 \times 10^{-8}$ s and $zv = 4.24$.

Relaxation time represents the duration required for spins or spin clusters to undergo flipping or rearrangement, and it is influenced by the associated spin-spin correlation length and the involved nature of magnetic interactions [45]. In canonical spin glasses, the randomly frozen spins have a very short spin-flip time, approximately on the order of atomic spin flip ($\tau_o \sim 10^{-13}$ s) [43,46]. Superparamagnets, consisting of small magnetic domains with short-range interactions, display a slightly longer $\tau_o$ ranging from $10^{-9}$ to $10^{-11}$s. Cluster glasses, formed by interacting spin clusters, have $\tau_o$ values between $10^{-8}$ and $10^{-12}$ s. Systems with single-molecule magnets exhibit $\tau_o$ of approximately $10^{-6}$ s, while skyrmions or antiskyrmions, characterized by larger size or topological features, generally demonstrate even longer relaxation times ($\tau_o \sim 10^{-5}$- $10^{-2}$ s) [43,45,47]. Generally, as the effective mass of the magnetic entity increases, where spins collectively respond to external perturbations, the spin flip time $\tau_o$ progressively lengthens, depending on the effective size or strength of interactions within the cluster. Given that the current system does not represent a chemically segregated phase, as evidenced by compositional (see SI Figure 2, [19]), $\chi'(T)$ analyses, and the $M(T)$ as well as $M(H)$ measurements, which exclude the possibility of any superparamagnetic state, the field-induced first-order, glassy magnetic phase with $\tau_o \sim 10^{-8}$ s indicates the formation of spin clusters or nanoscale spin textures in GdTi$_3$Bi$_4$ around $H_{C2} \sim 3.4$ T.

The magnetoresistance (MR%) at different temperatures is shown in Figure 3(a), calculated as $MR\% = [\rho_{xx}(H) - \rho_{xx}(0)] \times 100/\rho_{xx}(0)$. At low temperatures ($T < T_N = 14.2$ K), MR% exhibits an initial semiquadratic increase in the low-field regime (0 to 1.7 T), reaching ~5% at 1.6 T and 2 K. This behaviour is likely due to electron-hole compensation, [48] as indicated by the presence of both electron and hole charge carriers in the Hall analysis. At higher fields, MR% initially shows a positive jump at $H_{C1} \sim 1.7$ T, followed by a negative jump at $H_{C2} \sim 3.4$ T. These fields correspond to the first and second magnetization jumps as shown in Figure 2(b). At $H_{C1}$ the MR% increases by 26.6% while at $H_{C2}$ the MR% decreases by 27.7%. The synchronized changes in MR% and magnetization underscore the critical role of magnetic moment reorientation between adjacent layers of this antiferromagnetic system. At $H_{C1}$, the onset of first magnetization plateau likely leads to the formation of high-resistance state, characterized by adjacent magnetic layers with opposite moment orientation [25] (see SI Figures S11(a, b) [19]). At $H_{C2}$, the second magnetization jump predominantly align moments within the magnetic layers in the same direction, resulting to a low-resistance state. This gives rise to jumps in MR similar to those observed in giant magnetoresistance multilayers [26]. In the intermediate field regions (for $B > H_{C2}$), the negative magnetoresistance resulting from the suppression of spin disorder scattering is in competition with the positive magnetoresistance induced by the Lorentz force [49]. At temperatures exceeding $T_N$, such as 15 K, a significant negative MR% ($\sim 23.4\%$) is primarily attributed to the suppression of spin disorder scattering [50].

The Hall resistivity of GdTi₃Bi₄, depicted in Figure 3(b), exhibits complexity due to the combined effects of multiple band contributions, magnetization jumps or MMT, and the formation of spin textures. The Hall resistivity is non-linear, displaying both positive and negative slopes, which suggests contributions from both electrons and holes. The observed jumps in Hall resistivity at critical fields $H_{C1}$ and $H_{C2}$ correspond with the magnetization jumps, thereby demonstrating the anomalous Hall effect. The field-dependent Hall conductivity, calculated using $\sigma_{xy} = \rho_{yx}/(\rho_{yx}^2 + \rho_{xx}^2)$, is illustrated in Figure 3(c). The pronounced jumps at the MMT indicate a giant anomalous Hall effect (AHE) in GdTi₃Bi₄.

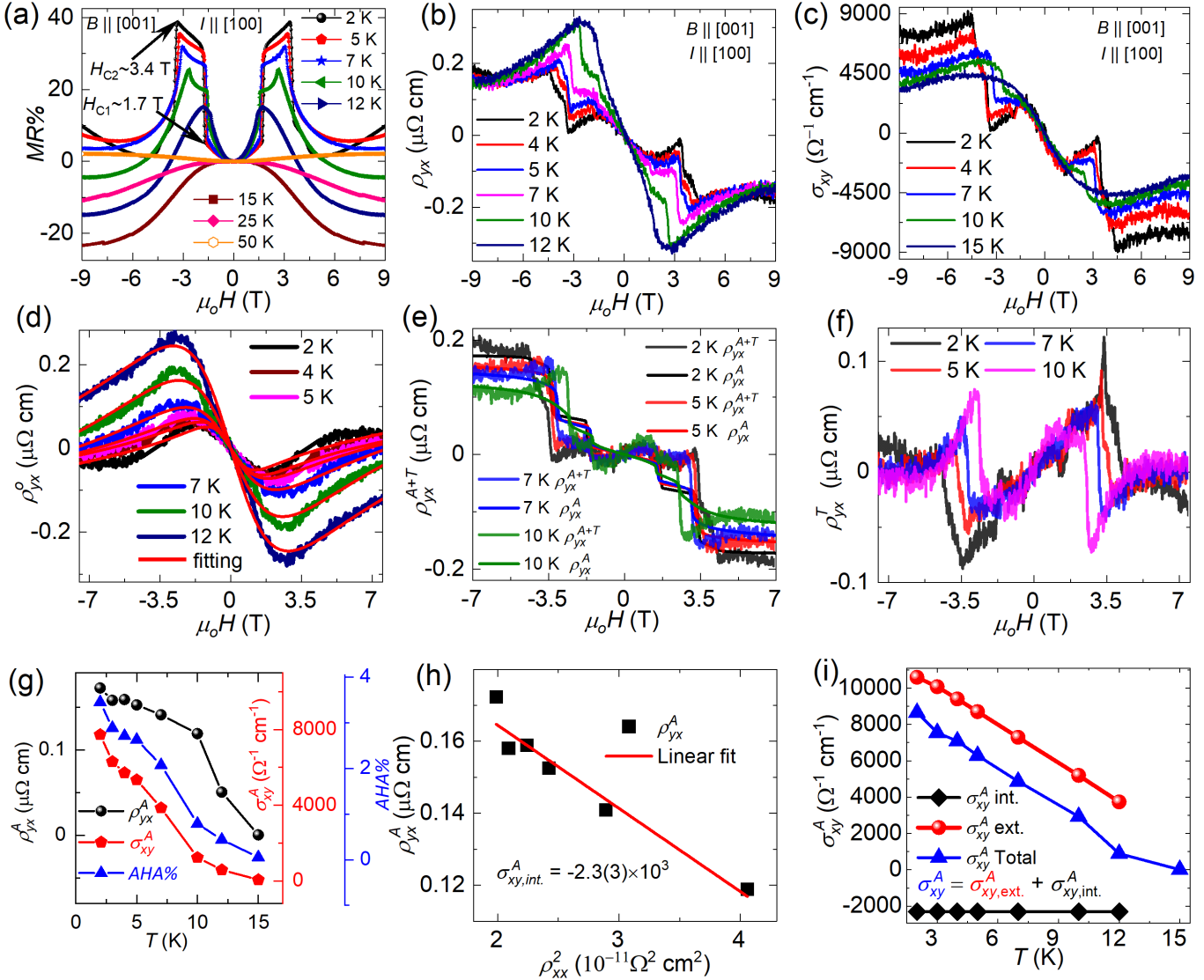

**Figure 3.** Magnetic field dependence of (a) MR%, (b) Hall resistivity ($\rho_{yx}$), and (c) Hall conductivity ($\sigma_{xy}$), at different temperatures. The magnetic field ($H$) is applied along [001] and current ($I$) is applied along [100]. (d) Nonlinear ordinary Hall resistivity ($\rho_{yx}^o$) fitted using two-band model. (e) Anomalous and spin texture induced additional Hall resistivity ($\rho_{yx}^{A+T}$) along with simulated anomalous Hall resistivity ($\rho_{yx}^A$), (f) Field dependent additional Hall resistivity ($\rho_{yx}^T$) at different temperatures. (g) Temperature dependence of $\rho_{yx}$, anomalous Hall conductivity ($\rho_{yx}^A$), and anomalous Hall angle (AHA%). (h) Scaling of the anomalous Hall resistivity using Tian-Ye-Jin (TYJ) model. (i) Temperature dependence of total anomalous Hall conductivity ($\sigma_{xy}^A$), extrinsic anomalous Hall conductivity ($\sigma_{xy,ex}^A$) and intrinsic anomalous Hall conductivity ($\sigma_{xy,in}^A$).

According to Matthiessen's rule the total Hall resistivity is given by:

$$\rho_{yx} = \rho_{yx}^o + \rho_{yx}^A + \rho_{yx}^T \tag{1}$$

Where $\rho_{yx}^o$ is ordinary Hall resistivity due to Lorentz force, $\rho_{yx}^A$ is anomalous Hall resistivity from magnetization and $\rho_{yx}^T$ represents additional Hall resistivity arising from the so-called topological Hall contributions from trivial or non-trivial spin textures etc. To isolate the ordinary Hall component, first the sudden jumps in $\rho_{yx}$ caused by MMT were subtracted by extrapolating the ordinary data near these points where magnetization shows no discontinuities (detailed procedure for extracting $\rho_{yx}^o$, $\rho_{yx}^A$, $\rho_{yx}^T$ is shown in SI Figure S12 (a-i) [19]). The data was then fitted with the two-band model given as

$$\rho_{yx}^o = \frac{H}{e} \frac{(n_h\mu_h^2 - n_e\mu_e^2) + (n_h - n_e)\mu_h^2\mu_e^2 H^2}{(n_h\mu_h^2 + n_e\mu_e^2)^2 + (n_h - n_e)^2\mu_h^2\mu_e^2 H^2} \tag{2}$$

The fitting is illustrated in Figure 3(d). At 2 K, the estimated high electron mobility ($\mu_e$ = 7435 cm$^2$ V$^{-1}$ s$^{-1}$) and low carrier density ($n_e$ = 5×10$^{18}$ cm$^{-3}$) (see SI Figures S13(a, b) [19]), suggests a small electron pocket possibly from linear bands near $E_F$ [13], aligning well with the band structure calculations in SI Figure S7(b, c) [19].

The combined anomalous and additional Hall resistivity $\rho_{yx}^{A+T} = \rho_{yx}^A + \rho_{yx}^T$ is determined by subtracting the $\rho_{yx}^o$ from $\rho_{yx}$ as illustrated in Figure 3(e). The anomalous Hall resistivity is expressed as $\rho_{yx}^A = 4\pi R_S M$. By analysing $\rho_{yx}^{A+T}$ in the high field regime, where spin textures or magnetization saturates and $\rho_{yx}^T$ vanishes, the intercept yields $4\pi R_s$ [51]. The simulated anomalous Hall resistivity ($4\pi R_s M$) is also depicted in Figure 3(e). The extracted additional Hall resistivity is presented in Figure 3(f) (contour plot of additional Hall resistivity as a function of temperature and magnetic field is shown in SI Figure S14 [19]). At 2 K, $\rho_{yx}^T$ attains a maximum value of up to 0.12 $\mu\Omega$ cm, which is significantly larger than $\rho_{yx}^T$ in well studied skyrmionic systems like MnSi (0.004 $\mu\Omega$ cm), MnP (0.01$\mu\Omega$ cm), and other comparable systems like MnGe (0.16 $\mu\Omega$ cm) though smaller than in EuO (6 $\mu\Omega$ cm) and correlated oxides (200 $\mu\Omega$ cm) [51]. This substantial additional Hall signal, in conjunction with the first-order phase transition in dc magnetization and the frequency-dependent ac susceptibility peak shift with a relaxation time of approximately 10$^{-8}$ s, clearly indicates the formation of field-induced nanoscale spin textures in the system. Recent investigations into domain morphologies using magnetic force microscopy (MFM) conducted on GdTi$_3$Bi$_4$ single crystals corroborate these findings. E. Cheng *et al.* [13] observed a maximum density of striped magnetic domains and domain walls in their MFM study around the MMT of 3.4 T field, while J. Guo *et al.* [16] noted randomized magnetic bubble domain orientations in similar field regions. These MFM studies indicate possibility of topological spin textures.

To analyse the anomalous Hall contribution, we concentrated on higher magnetic fields (above ~ 4.2 T), where magnetization reaches saturation. Above ~ 4.2 T, the resistivity contribution from the spin texture-induced additional Hall resistivity vanishes in $\rho_{yx}^{A+T}$ (Figure 3(e)). Consequently, the anomalous Hall resistivity at various temperatures can be determined by linearly extrapolating the high-field data to zero field. Figure 3(g) illustrates the temperature dependence of anomalous Hall resistivity, anomalous Hall conductivity, and anomalous Hall angle. At 2 K, GdTi$_3$Bi$_4$ exhibits a colossal anomalous Hall conductivity of $\sigma_{xy}^A = 8.6(7) \times 10^3$ $\Omega^{-1}$ cm$^{-1}$ with a substantial AHA% of 3.45 %. Further analysis of the temperature dependent $\sigma_{xy}^A$ suggests that extrinsic mechanisms predominantly influence the observed AHE. The intrinsic AHE contribution is generally temperature-independent, as the chemical potential does not significantly vary with temperature. Utilizing the Tian-Ye-Jin (TYJ) model, we have delineated the intrinsic and extrinsic contributions, as depicted in Figure 3(h) [28,52]. The substantial anomalous Hall conductivity at 2 K, $\sigma_{xy}^A = 8.6(7) \times 10^3$ $\Omega^{-1}$ cm$^{-1}$ comprises two components: skew scattering $\sigma_{xy,sk}^A = 1.06(4) \times 10^4$ $\Omega^{-1}$ cm$^{-1}$ attributable to extrinsic effects, and intrinsic anomalous Hall conductivity $\sigma_{xy,in}^A = -2.3(3) \times 10^3$ $\Omega^{-1}$ cm$^{-1}$ arising from the Berry curvature of the occupied electronic states (see SI Figure S15 (a-d) [19]). Notably, both components exhibit opposing behaviours, consistent with observations in numerous prior studies [31,53–56]. Other extrinsic effect like side jump contributions, given by $\sigma_{xy,sj}^A = \frac{e^2}{hc}\left(\frac{E_{SO}}{E_F}\right)$, where $E_{SO}$ represents spin-orbit interaction energy, and $E_F$ denotes the Fermi energy, and $c$ is lattice parameter, are typically minimal in spin-polarized ferromagnetic states, as $\frac{E_{SO}}{E_F}$ is generally less than 0.01 [57]. The temperature dependence of the total ($\sigma_{xy}$) and extrinsic anomalous Hall conductivity $\sigma_{xy,ex}^A$, as shown in Figure 3(i), decreases with increasing temperature.

The various sources contributing to Hall conductivity are more effectively analysed using the universal plot of anomalous Hall conductivity, as depicted in Figure 4(a). In this context, AHC of GdTi$_3$Bi$_4$ is compared with that of several other well-established systems [2,3,31,58–65]. Based on longitudinal conductivity or scattering time scales, experimental Hall systems typically fall into three scattering regimes. The universal plot positions GdTi$_3$Bi$_4$ within the intrinsic regime, near the boundary of the skew scattering regime. This positioning suggests that main mechanism behind large AHC in GdTi$_3$Bi$_4$ are intrinsic Berry curvature effects and skew scattering, while side jump contributions playing minimal role. The substantial skew scattering contribution may have arisen from the asymmetric scattering of electrons due to the combined effect of residual impurities and field-induced spin clusters as observed in the magnetisation experiments. To elucidate the origin of the large intrinsic AHC of GdTi$_3$Bi$_4$, the energy dependence of Hall conductivity is theoretically calculated, as shown in Figure 4(b). Notably, at the Fermi level, the theoretically calculated intrinsic AHC ($\sigma_{xy,in}^A$) is approximately 500 $\Omega^{-1}$ cm$^{-1}$, which is significantly smaller compared to the experimentally observed $\sigma_{xy,in}^A \approx$ -2.3(3)×10$^3$ $\Omega^{-1}$ cm$^{-1}$ at 2 K and in agreement with recent report [80]. In GdTi$_3$Bi$_4$, the bands at the Fermi level are dominated by the Ti atoms, which form a kagome lattice, a structure that naturally gives rise to flat bands. Approximately 0.5 eV above the Fermi energy, the strong coupling of flat bands results in an AHC of ∼ 1300 $\Omega^{-1}$ cm$^{-1}$. Although the calculated value remains lower than the experimentally observed one, we assert that this discrepancy may arise from two potential sources: (i) the limitations of density functional theory (DFT) calculations in accurately describing $f$-electrons, or (ii) the inadequacy of our first-principles calculations in properly accounting for the magnetic structure of GdTi$_3$Bi$_4$ under experimental conditions. As the magnetic field increases, GdTi$_3$Bi$_4$ undergoes a magnetic transition from an AFM structure to a possible non-collinear state, eventually reaching a FM state. However, the transition state between the initial AFM and the FM is too complex for DFT to capture, and the calculations only yield results from the final collinear FM states. Consequently, contributions from the nonlinear magnetic phase may be omitted in the DFT calculations. However, the spiking trend of AHC at higher energies in Figure 4(b) suggests that the $f$-electrons can substantially enhance the AHE, underscoring the possible role of $f$-electrons in augmenting the AHE in GdTi$_3$Bi$_4$. Additionally, the Coulomb interaction parameter $U$ in Gd was varied between 2 - 6 eV to assess its impact on the electronic band structure and AHC. As demonstrated in SI Figure S16-S17 [19], increasing $U$ shifts the flat bands away from the Fermi level, resulting in a reduced AHC.

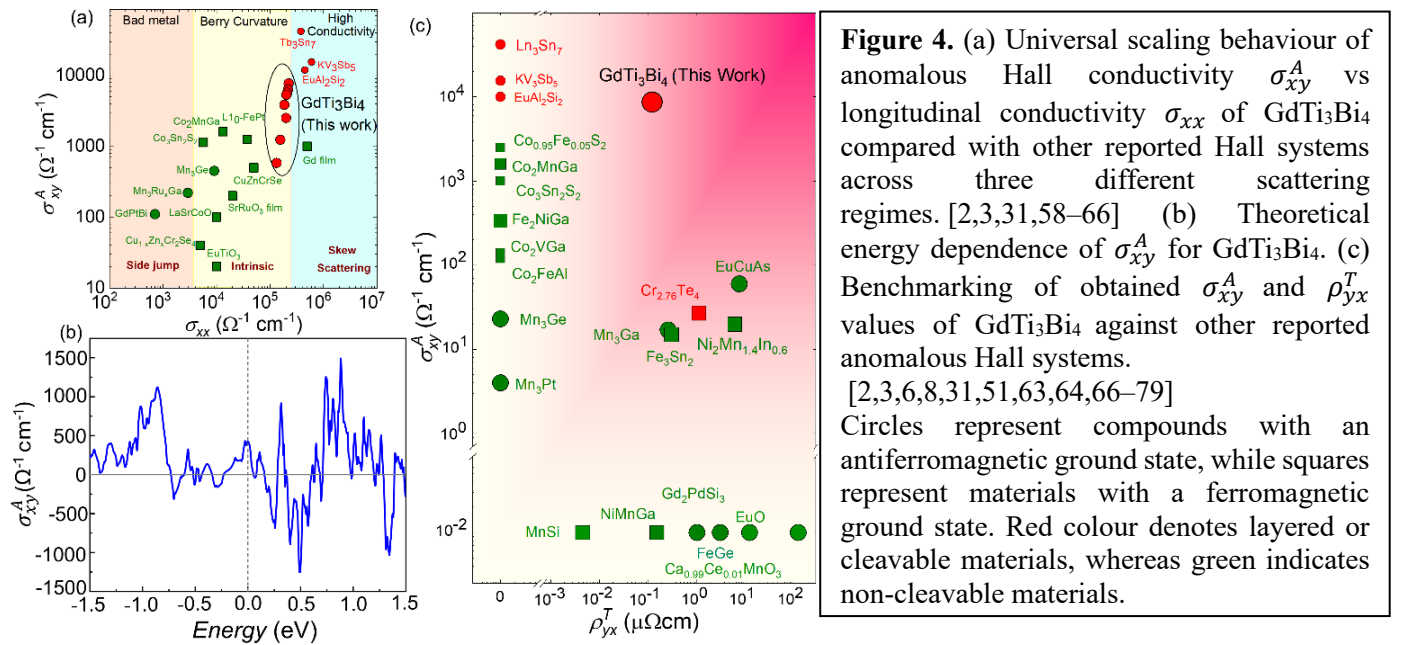

**Figure 4.** (a) Universal scaling behaviour of anomalous Hall conductivity $\sigma_{xy}^A$ vs longitudinal conductivity $\sigma_{xx}$ of GdTi$_3$Bi$_4$ compared with other reported Hall systems across three different scattering regimes. [2,3,31,58–66] (b) Theoretical energy dependence of $\sigma_{xy}^A$ for GdTi$_3$Bi$_4$. (c) Benchmarking of obtained $\sigma_{xy}^A$ and $\rho_{yx}^T$ values of GdTi$_3$Bi$_4$ against other reported anomalous Hall systems. [2,3,6,8,31,51,63,64,66–79] Circles represent compounds with an antiferromagnetic ground state, while squares represent materials with a ferromagnetic ground state. Red colour denotes layered or cleavable materials, whereas green indicates non-cleavable materials.

The giant anomalous Hall conductivity $\sigma_{xy}^A$ and the additional Hall resistivity $\rho_{yx}^T$ of GdTi$_3$Bi$_4$ is compared with other known compounds exhibiting similar effects as illustrated in Figure 4(c). GdTi$_3$Bi$_4$ is a unique system that simultaneously demonstrates dual Hall conductivity $\sigma_{xy}^A = 8.6(7)\times10^3\,\Omega^{-1}$ cm$^{-1}$ (anomalous) and spin-texture driven additional Hall resistivity $\rho_{yx}^T = 0.12\ \mu\Omega$ cm (spin texture) at 2 K and these values are significant for a layered kagome antiferromagnetic system. These findings establish GdTi$_3$Bi$_4$ as a rare platform hosting both giant anomalous and field-induced additional Hall responses, thereby opening new avenues for the design of materials with significant Hall effects and potential inter-controllability of the real-space and momentum-space Berry curvature effects.

In summary, $GdTi_3Bi_4$ is demonstrated as a unique layered kagome antiferromagnetic system, that hosts a remarkable combination of large anomalous and spin texture related Hall responses. Additionally, the system exhibits multiple field-induced metamagnetic states. The presence of a first-order phase transition, along with a frequency-dependent shift in the ac susceptibility peak and a pronounced spin texture Hall effect, suggests the formation of a non-equilibrium, glassy magnetic state with slow spin dynamics, potentially indicative of a field-induced spin-texture phase in $GdTi_3Bi_4$. The significant spin texture Hall effect (up to 0.12 $\mu\Omega$ cm) combined with a giant anomalous Hall conductivity of $8.6(7)\times10^3$ $\Omega^{-1}$ cm$^{-1}$ at 2 K establishes new benchmarks for layered magnetic systems. Although the precise microscopic spin textures remain to be directly resolved, the transport anomalies, scaling analyses, and Berry-curvature calculations collectively indicate a strong interplay between the kagome-lattice electronic structure and anisotropic magnetism in $GdTi_3Bi_4$. The identification of such a layered material contributes to the expanding family of functional kagome magnets and underscores the potential for exfoliation, external control, and device integration. Consequently, our investigation opens new avenues for the development of smart materials with exotic tunable Hall effects for next-generation spintronic, spin-orbit torque, and Hall sensing applications.

## Acknowledgements

KM acknowledge Max Planck Society for the funding support under Max Planck-India partner group project; Board of Research in Nuclear Sciences (BRNS) under grant no: 58/20/03/2021-BRNS/37084/DAE-YSRA; Science and Engineering Research Board, DST, Government of India, via grant no: CRG/2022/001826; Aeronautics Research and Development Board (ARDB, Project No. 1992); Defence Research and Development Organization (DRDO) with project no: DFTM/033203/P/41/JATC-P2QP-17/10/D(R&D)/2023. SS thanks MHRD for Institute fellowship and funding support to GIMRT Program of the Institute for Materials Research, Tohoku University (Proposal No. 202408-CRKEQ-0510). SR acknowledges ANRF India for the National Postdoctoral Fellowship (Grant No. PDF/2025/003558). Authors also thank Central research facility (CRF), and Nanoscale research facility at IIT Delhi for providing materials characterization facility, like PPMS, MPMS, EDX, XRD etc.

## Conflict of Interest:

The authors declare no conflict of interest.

## Reference

[1]    R. Karplus and J. M. Luttinger, Hall Effect in Ferromagnetics, Phys. Rev. **95**, 1154 (1954).

[2]     E. Liu, Y. Sun, N. Kumar, L. Muechler, A. Sun, L. Jiao, SY. Yang, A. Lang, Q. Xu, J. Kroder et al., Giant anomalous Hall effect in a ferromagnetic kagome-lattice semimetal, Nat. Phys. **14**, 1125 (2018).

[3]     K. Manna, L. Muechler, T. H. Kao, R. Stinshoff, Y. Zhang, J. Gooth, N. Kumar, From Colossal to Zero: Controlling the Anomalous Hall Effect in Magnetic Heusler Compounds via Berry Curvature Design, Phys. Rev. X **8**, 041045 (2018).

[4]     N. Kanazawa, Y. Onose, T. Arima, D. Okuyama, K. Ohoyama, S. Wakimoto, K. Kakurai, S. Ishiwata, and Y. Tokura, Large Topological Hall Effect in a Short-Period Helimagnet MnGe, Phys. Rev. Lett. **106**, 156603 (2011).

[5]     S. Mühlbauer, B. Binz, F. Jonietz, C. Pfleiderer, A. Rosch, A. Neubauer, R. Georgii, and P. Böni, Skyrmion Lattice in a Chiral Magnet, Science **323**, 915 (2009).

[6]     T. Kurumaji, T. Nakajima, M. Hirschberger, A. Kikkawa, Y. Yamasaki, H. Sagayama, H. Nakao, Y. Taguchi, T. Arima, and Y. Tokura, Skyrmion lattice with a giant topological Hall effect in a frustrated triangular-lattice magnet, Science **365**, 914 (2019).

[7]     S. X. Huang and C. L. Chien, Extended Skyrmion Phase in Epitaxial FeGe (111) Thin Films, Phys. Rev. Lett. **108**, 267201 (2012).

[8]     A. Neubauer, C. Pfleiderer, B. Binz, A. Rosch, R. Ritz, P. G. Niklowitz, and P. Böni, Topological Hall Effect in the A Phase of MnSi, Phys. Rev. Lett. **102**, 186602 (2009).

[9]     N. Nagaosa and Y. Tokura, Topological properties and dynamics of magnetic skyrmions, Nat. Nanotechnol. **8**, 899 (2013).

[10]    J. Guo, L. Zhou, J. Ding, G. Qu, Z. Liu, Y. Du, H. Zhang, J. Li, Y. Zhnag, F. Zhou et al., Tunable magnetism and band structure in kagome materials RETi3Bi4 family with weak interlayer interactions, Sci. Bull. (Beijing) **69**, 2660 (2024).

[11]    T. Jungwirth, X. Marti, P. Wadley, and J. Wunderlich, Antiferromagnetic spintronics, Nat. Nanotechnol. **11**, 231 (2016).

[12]    B. Huang, M. A. McGuire, A. F. May, D. Xiao, P. Jarillo-Herrero, and X. Xu, Emergent phenomena and proximity effects in two-dimensional magnets and heterostructures, Nat. Mater. **19**, 1276 (2020).

[13]    E. Cheng, N. Mao, X. Yang, B. Song, R. Lou, T. Ying, S. Nie, A. Fedorov, F. Bertran, P. Ding et al., Striped magnetization plateau and chirality-reversible anomalous Hall effect in a magnetic kagome metal, arXiv:2409.01365 (2024).

[14]    B. R. Ortiz, H. Miao, DS. Parker, F. Yang, GD. Samolyuk, EM. Clements, A. Rajapitamahuni, T. Yilmaz, E. Vescovo, J. Yan, AF. May, MA. McGuire, Evolution of Highly Anisotropic Magnetism in the Titanium-Based Kagome Metals LnTi3Bi4 (Ln: La···Gd3+, Eu2+, Yb2+), Chem. Mater. **35**, 9756 (2023).

[15] X. Li, Y. Yang, F. Guan, X. Zhu, W. Ning, and M. Tian, Anisotropic magnetoresistance in antiferromagnetic kagome metal GdTi3Bi4, Appl. Phys. Lett. **126**, 092405 (2025).

[16] J. Guo, S. Zhu, R. Zhou, R. Wang, Y. Wang, J. Sun, Z. Zhao, X. Dong. J. Cheng, Tunable Bifurcation of Magnetic Anisotropy and Bi-Oriented Antiferromagnetic Order in Kagome Metal GdTi3Bi4, Phys. Rev. Lett. **134**, 226704 (2025).

[17] X. Yang, J. Pan, M. Yang, D. Chu, and S. Liu, Two-dimensional Critical behavior, complex magnetic phase diagram, and anisotropic magnetic entropy change of a two-dimensional kagome magnet GdTi3Bi4, J. Magn. Magn. Mater. **621**, 172916 (2025).

[18] X. Han, H. Chen, Z. Cao, J. Guo, F. Fei, H. Tan, J. Guo, Y. Shi, R. Zhou, R. Wang, Z. Zhao, H. Yang, et al., Discovery of Unconventional Charge-Spin-Intertwined Density Wave in Magnetic Kagome Metal GdTi3Bi4, arXiv: 2503.05545 (2025).

[19] See Supplementary Information at [URL] for sample synthesis, structural and compositional analysis, details and analysis of magnetization and electrical transport measurements along different crystallographic directions, and theoretical calculations. The Supplementary Information also contains Refs. [20-31].

[20] G. Kresse and J. Hafner, *Ab initio* molecular dynamics for liquid metals, Phys. Rev. B **47**, 558 (1993).

[21] G. Kresse and J. Furthmüller, Efficient iterative schemes for *ab initio* total-energy calculations using a plane-wave basis set, Phys. Rev. B **54**, 11169 (1996).

[22] J. P. Perdew, K. Burke, and M. Ernzerhof, Generalized Gradient Approximation Made Simple, Phys. Rev. Lett. 77, 3865 (1996).

[23] J. R. Yates, X. Wang, D. Vanderbilt, and I. Souza, Spectral and Fermi surface properties from Wannier interpolation, Phys. Rev. B **75**, 195121 (2007).

[24] N. Nagaosa, J. Sinova, S. Onoda, A. H. MacDonald, and N. P. Ong, Anomalous Hall effect, Rev. Mod. Phys. **82**, 1539 (2010).

[25] E. J. Telford, A.H. Dismukes, K. Lee, M. Cheng, A. Wieteska, A.K. Bartholomew, Y.S.Chen, X. Xu, A.N. Pasupathy, X. Zhu, C.R. Dean, X. Roy Layered Antiferromagnetism Induces Large Negative Magnetoresistance in the van der Waals Semiconductor CrSBr, Adv. Mater. **32**, 2003240 (2020).

[26] H. Yamamoto, Y. Motomura, T. Anno, and T. Shinjo, Magnetoresistance of non-coupled [NiFe/Cu/Co/Cu] multilayers, J. Magn. Magn. Mater. 126, 437 (1993).

[27] S. Albarakati, C. Tan, Z.J. Chen, J.G. Partridge, G. Zheng, L. Farrar, E.L.H. Mayes, M.R. Field, C. Lee, Y. Wang, Y. Xiong, M. Tian, F. Xiang, A.R. Hamilton et al., Antisymmetric magnetoresistance in van der Waals Fe3GeTe2 /graphite/Fe3GeTe2 trilayer heterostructures, Sci. Adv. **5**, aaw0409 (2019).

[28]  D. Hou, G. Su, Y. Tian, X. Jin, S. A. Yang, and Q. Niu, Multivariable Scaling for the Anomalous Hall Effect, Phys. Rev. Lett. 114, 217203 (2015).

[29]  L. Ye, Y. Tian, X. Jin, and D. Xiao, Temperature dependence of the intrinsic anomalous Hall effect in nickel, Phys. Rev. B 85, 220403 (2012).

[30]  H. Lv, X. C. Huang, K. H. L. Zhang, O. Bierwagen, and M. Ramsteiner, Underlying Mechanisms and Tunability of the Anomalous Hall Effect in NiCo2O4 Films with Robust Perpendicular Magnetic Anisotropy, Adv. Sci. **10**, 2302956 (2023).

[31]  J. Chen, X. Yang, F. Zhou, Y.-C. Lau, W. Feng, Y. Yao, Y. Li, Y. Jiang, and W. Wang, Colossal anomalous Hall effect in the layered antiferromagnetic EuAl2Si2 compound, Mater. Horiz. **11**, 4665 (2024).

[32]  D. Ram, J. Singh, M. K. Hooda, K. Singh, V. Kanchana, D. Kaczorowski, and Z. Hossain, Multiple magnetic transitions, metamagnetism, and large magnetoresistance in GdAuGe single crystals, Phys. Rev. B **108**, 235107 (2023).

[33]  D. C. Johnston, Magnetic Susceptibility of Collinear and Noncollinear Heisenberg Antiferromagnets, Phys. Rev. Lett. **109**, 077201 (2012).

[34]  S. Rathod, M. Malasi, A. Lakhani, and D. Kumar, Electron-magnon scattering in an anisotropic half-metallic ferromagnetic Weyl semimetal Co3Sn2S2, Phys. Rev. Mater. **6**, 084202 (2022).

[35]  S. Manni, S. L. Bud'ko, and P. C. Canfield, GdPtPb: A noncollinear antiferromagnet with distorted kagome lattice, Phys. Rev. B **96**, 54435 (2017).

[36]  S. K. Chhetri, G. Acharya, D. Graf, R. Basnet, S. Rahman, M.M. Sharma, D. Upreti, M.R.U. Nabi, S. Kryvyi, J. Sakon, M. Mortazavi, B. Da, H. Churchill, J. Hu, Large negative magnetoresistance in antiferromagnetic Gd2Se3, Phys. Rev. B **111**, 014431 (2025).

[37]  M. Lee, E. S. Choi, X. Huang, J. Ma, C. R. Dela Cruz, M. Matsuda, W. Tian, Z. L. Dun, S. Dong, and H. D. Zhou, Magnetic phase diagram and multiferroicity of Ba3MnNb2O9: A spin- 5/2 triangular lattice antiferromagnet with weak easy-axis anisotropy, Phys. Rev. B **90**, 224402 (2014).

[38]  Y. Okamoto, M. Tokunaga, H. Yoshida, A. Matsuo, K. Kindo, and Z. Hiroi, Magnetization plateaus of the spin-1/2 kagome antiferromagnets volborthite and vesignieite, Phys. Rev. B **83**, 180407 (2011).

[39]  M. A. Manekar, S. Chaudhary, M. K. Chattopadhyay, K. J. Singh, S. B. Roy, and P. Chaddah, First-order transition from antiferromagnetism to ferromagnetism in Ce(Fe0.96Al0.04)2, Phys. Rev. B **64**, 104416 (2001).

[40]  R. Das and P. Poddar, Observation of exchange bias below incommensurate antiferromagnetic (ICAFM) to canted A-type antiferromagnetic (cAAFM) transition in nanocrystalline orthorhombic EuMnO3, RSC Adv. **4**, 10614 (2014).

[41] A. Banerjee, A. K. Pramanik, K. Kumar, and P. Chaddah, Coexisting tunable fractions of glassy and equilibrium long-range-order phases in manganites, J. Phys.: Condens. Matter. **18**, L605 (2006).

[42] K. Kumar, A. K. Pramanik, A. Banerjee, P. Chaddah, S. B. Roy, S. Park, C. L. Zhang, and S. W. Cheong, Relating supercooling and glass-like arrest of kinetics for phase separated systems: Doped Ce Fe2 and (La,Pr,Ca) Mn O3, Phys. Rev. B Condens. Matter Mater. Phys. **73**, 184435 (2006).

[43] K. Manna, D. Samal, A. K. Bera, S. Elizabeth, S. M. Yusuf, and P. S. Anil Kumar, Correspondence between neutron depolarization and higher order magnetic susceptibility to investigate ferromagnetic clusters in phase separated systems, J. Phys.: Condens. Matter. **26**, 016002 (2014).

[44] V. K. Anand, D. T. Adroja, and A. D. Hillier, Ferromagnetic cluster spin-glass behavior in PrRhSn3, Phys. Rev. B **85**, 014418 (2012).

[45] J. A. Mydosh, Spin Glass: An Experimental Introduction, Taylor and Francis, London, 1993.

[46] K. Manna, D. Samal, S. Elizabeth, H. L. Bhat, and P. S. Anil Kumar, On the Magnetic Ground State of La0.85Sr0.15CoO3 Single Crystals, J. Phys. Chem. C **115**, 13985 (2011).

[47] P. V. P. Madduri, S. Sen, B. Giri, D. Chakrabartty, S. K. Manna, S. S. P. Parkin, and A. K. Nayak, ac susceptibility study of magnetic relaxation phenomena in the antiskyrmion-hosting tetragonal Mn-Pt(Pd)-Sn system, Phys. Rev. B **102**, 174402 (2020).

[48] S. Rathod, M. Malasi, A. Lakhani, and D. Kumar, How to Enhance Anomalous Hall Effects in Magnetic Weyl Semimetal Co3Sn2S2, Chem. Mater. **36**, 7418 (2024).

[49] N. Kumar, Y. Soh, Y. Wang, and Y. Xiong, Magnetotransport as a diagnostic of spin reorientation: Kagome ferromagnet as a case study, Phys. Rev. B **100**, 214420 (2019).

[50] K. Usami, Magnetoresistance in Antiferromagnetic Metals, J. Phys. Soc. Jpn. **45**, 466 (1978).

[51] L. Vistoli, W. Wang, A. Sander, Q. Zhu, B. Casals, R. Cichelero, A. Barthelemy, S. Fusil, G. Herranz, S. Valencia, R. Abrudan, et al., Giant topological Hall effect in correlated oxide thin films, Nat. Phys. **15**, 67 (2019).

[52] L. Ye, Y. Tian, X. Jin, and D. Xiao, Temperature dependence of the intrinsic anomalous Hall effect in nickel, Phys. Rev. B **85**, 220403 (2012).

[53] W. Xia, B. Bai, X. Chen, Y. Yang, Y. Zhang, J. Yuan, Q. Li, K. Yang, Z. Liu, Y. Shi, H. Ma, Y. Yang, M. He, et al., Giant Domain Wall Anomalous Hall Effect in a Layered Antiferromagnet EuAl2Si2, Phys. Rev. Lett. **133**, 216602 (2024).

[54] L. Wu, Y. Li, J. Xu, D. Hou, and X. Jin, Anisotropic intrinsic anomalous Hall effect in epitaxial Fe films on GaAs (111), Phys. Rev. B **87**, 155307 (2013).

[55] J. Gong, H. Wang, K. Han, X.-Y. Zeng, X.-P. Ma, Y.-T. Wang, J.-F. Lin, X.-Y. Wang, and T.-L. Xia, Anomalous Hall effect in an antiferromagnetic CeGaSi single crystal, Phys. Rev. B **109**, 024434 (2024).

[56] W. J. Fan, L. Ma, and S. M. Zhou, Sign change of skew scattering induced anomalous Hall conductivity in epitaxial NiCo(002) films: band filling effect, J. Phys. D: Appl. Phys. **48**, 195004 (2015).

[57] S. Onoda, N. Sugimoto, and N. Nagaosa, Intrinsic Versus Extrinsic Anomalous Hall Effect in Ferromagnets, Phys. Rev. Lett. **97**, 126602 (2006).

[58] A. K. Nayak, J. K. Fischer, Y. Sun, B. Yan, J. Karel, A.C. Komarek, C. Shekhar, N. Kumar, W. Schnelle, J. Kubler, C. Felser, S.S.P. Parkin, Large anomalous Hall effect driven by a nonvanishing Berry curvature in the noncolinear antiferromagnet Mn3Ge, Sci. Adv. **2**, e1501870 (2016).

[59] J. Yu, U. Ruediger, A. D. Kent, R. F. C. Farrow, R. F. Marks, D. Weller, L. Folks, and S. S. P. Parkin, Magnetotransport and magnetic properties of molecular-beam epitaxy L1 FePt thin films, J. Appl. Phys. **87**, 6854 (2000).

[60] N. Thiyagarajah, Y.-C. Lau, D. Betto, K. Borisov, J. M. D. Coey, P. Stamenov, and K. Rode, Giant spontaneous Hall effect in zero-moment Mn2RuxGa, Appl. Phys. Lett. **106**, 122402 (2015).

[61] G. Skorupskii, F. Orlandi, I. Robredo, M. Jovanovic, R. Yamada, F. Katmer, M. G. Vergniory, P. Manuel, M. Hirschberger, and L. M. Schoop, Designing giant Hall response in layered topological semimetals, Nat. Commun. **15**, 10112 (2024).

[62] T. Miyasato, N. Abe, T. Fujii, A. Asamitsu, S. Onoda, Y. Onose, N. Nagaosa, and Y. Tokura, Crossover Behavior of the Anomalous Hall Effect and Anomalous Nernst Effect in Itinerant Ferromagnets, Phys. Rev. Lett. **99**, 086602 (2007).

[63] S.-Y. Yang, Y. Wang, B.R. Ortiz, D. Liu, J. Gayles, E. Derunova, R.G. Hernandez, L. Smejkal, Y. Chen, S.S.P. Parkin, et al., Giant, unconventional anomalous Hall effect in the metallic frustrated magnet candidate, KV3Sb5, Sci. Adv. **6**, 6003 (2020).

[64] G. Skorupskii, F. Orlandi, I. Robredo, M. Jovanovic, R. Yamada, F. Katmer, M. G. Vergniory, P. Manuel, M. Hirschberger, and L. M. Schoop, Designing giant Hall response in layered topological semimetals, Nat. Commun. **15**, 10112 (2024).

[65] K. S. Takahashi, H. Ishizuka, T. Murata, Q. Y. Wang, Y. Tokura, N. Nagaosa, and M. Kawasaki, Anomalous Hall effect derived from multiple Weyl nodes in high-mobility EuTiO3 films, Sci. Adv. **4**, eaar7880 (2018).

[66] S. N. Guin, K. Manna, J. Noky, S.J. Watzman, C. Fu, N. Kumar, W. Schnelle, C. Shekhar, Y. Sun, J. Gooth, C. Felser, Anomalous Nernst effect beyond the magnetization scaling relation in the ferromagnetic Heusler compound Co2MnGa, NPG Asia Mater. **11**, 16 (2019).

[67] Y. Ohuchi, Y. Kozuka, M. Uchida, K. Ueno, A. Tsukazaki, and M. Kawasaki, Topological Hall effect in thin films of the Heisenberg ferromagnet EuO, Phys. Rev. B **91**, 245115 (2015).

[68] J. Choi, J. Park, W. Kyung, Y. Kim, M. K. Kim, J. Kwon, C. Kim, J. Rhim, S. Y. Park, and Y. Jo, Tunable Colossal Anomalous Hall Conductivity in Half-Metallic Material Induced by $d$-Wave-Like Spin-Orbit Gap, Adv. Sci. **11**, 2307288 (2024).

[69] I.-M. Imort, P. Thomas, G. Reiss, and A. Thomas, Anomalous Hall effect in the Co-based Heusler compounds Co2FeSi and Co2FeAl, J. Appl. Phys. **111**, 07D313 (2012).

[70] P. Yanda, L. Noohinejad, N. Mao, N. Peshcherenko, K. Imasato, A. K. Srivastava, Y. Guan, B. Giri, A. K. Sharma, K. Manna, et al., Giant Topological Hall Effect and Colossal Magnetoresistance in Heusler Ferromagnet near Room Temperature, Adv. Mater. **37**, 2411240 (2024).

[71] S. Purwar, A. Low, A. Bose, A. Narayan, and S. Thirupathaiah, Investigation of the anomalous and topological Hall effects in layered monoclinic ferromagnet Cr2.76Te4, Phys. Rev. Mater. **7**, 094204 (2023).

[72] S. Roychowdhury, K. Samanta, P. Yanda, B. Malaman, M. Yao, W. Schnelle, E. Guilmeau, P. Constantinou, S. Chandra, h. Borrmann, M.G. Vergniory, V. Strocov, C. Shekhar, C. Felser, Interplay between Magnetism and Topology: Large Topological Hall Effect in an Antiferromagnetic Topological Insulator, EuCuAs, J. Am. Chem. Soc. **145**, 12920 (2023).

[73] W. Zhang, B. Balasubramanian, A. Ullah, R. Pahari, X. Li, L. Yue, S. R. Valloppilly, A. Sokolov, R. Skomski, and D. J. Sellmyer, Comparative study of topological Hall effect and skyrmions in NiMnIn and NiMnGa, Appl. Phys. Lett. **115**, (2019).

[74] B. E. Zuniga-Cespedes, K. Manna, H. M. L. Noad, P.-Y. Yang, M. Nicklas, C. Felser, A. P. Mackenzie, and C. W. Hicks, Observation of an anomalous Hall effect in single-crystal Mn3Pt, New J. Phys. **25**, 023029 (2023).

[75] C. Wuttke, F. Caglieris, S. Sykora, F. Scaravaggi, A.U.B. Wolter, K. Manna, V. Suss, C. Shekhar, C. Felser, B. Buchner, C. Hess, Berry curvature unravelled by the anomalous Nernst effect in Mn3Ge, Phys. Rev. B **100**, 085111 (2019).

[76] J. C. Gallagher, K. Y. Meng, J. T. Brangham, H. L. Wang, B. D. Esser, D. W. McComb, and F. Y. Yang, Robust Zero-Field Skyrmion Formation in FeGe Epitaxial Thin Films, Phys Rev Lett **118**, 027201 (2017).

[77] F. Mende, J. Noky, S. N. Guin, G. H. Fecher, K. Manna, P. Adler, W. Schnelle, Y. Sun, C. Fu, and C. Felser, Large Anomalous Hall and Nernst Effects in High Curie-Temperature Iron-Based Heusler Compounds, Adv. Sci. **8**, 2100782 (2021).

[78] D. Zhang, Z. Hou, and W. Mi, Anomalous and topological Hall effects of ferromagnetic Fe3Sn2 epitaxial films with kagome lattice, Appl. Phys. Lett. **120**, (2022).

[79] S. Husain, A. Kumar, S. Akansel, P. Svedlindh, and S. Chaudhary, Anomalous Hall effect in ion-beam sputtered Co2FeAl full Heusler alloy thin films, J. Magn. Magn. Mater. 442, 288 (2017).

[80] A. K. Sharma, B. Tai, S. Roychowdhury, P. Yanda, U. Burkhardt, X. Feng, C. Felser, and C. Shekhar, Anisotropic anomalous Hall effect in distorted kagome $GdTi_3Bi_4$, Phys. Rev. B 113, L060402 (2026).

# Supporting Information

**Berry curvature induced giant anomalous and spin texture driven Hall responses in the layered kagome antiferromagnet GdTi$_3$Bi$_4$**

Shobha Singh,[1] Shivam Rathod,[1] Rong chen[2], Lipika[1], Sneh[1], Rie Y. Umetsu[3,4], Yan Sun[2] and Kaustuv Manna[1]*

[1]*Department of Physics, Indian Institute of Technology Delhi, New Delhi 110016, India*
[2]*Shenyang National Laboratory for Materials Science, Institute of Metal Research, Chinese Academy of Sciences 72 Wenhua Road, Shenyang 110016, China.*
[3]*Institute for Materials Research, Tohoku University, Sendai 980-8577, Japan*
[4]*Center for Science and Innovation in Spintronics, Tohoku University, Sendai 980-8577, Japan*

* E-mail: kaustuvmanna@iitd.ac.in

**Experimental Section**

*Single crystal growth.* The single crystals of GdTi$_3$Bi$_4$ were synthesized using the flux method, with bismuth serving as the flux. High-purity elements, Gd metal ingot (99.9 %), Ti powder (99.5 %), Bi (99.9 %) were measured in a molar ratio of 2:4:12 and placed in a pre-heated alumina crucible within an argon-filled glove box (O$_2$ ~ 0.1 ppm, H$_2$O ~ 0.1 ppm). The alumina crucible was subsequently inserted into a quartz tube, which was evacuated and sealed under a vacuum of 10$^{-3}$ mbar. The sealed quartz tube was positioned vertically in a muffle furnace, initially heated to 1050 °C at a rate of 100 °C/h, maintained at this temperature for 24 hours, and then subjected to slow cooling at a rate of 2 °C/h to 500 °C. At 500 °C, the ampule was placed in a high-temperature centrifuge for 4 minutes at 2000 rpm to remove the excess bismuth flux. The process yielded hexagonal, plate-like, shiny crystals, typically measuring 0.5 - 1.5 mm in width, up to 3 mm in length, and up to 0.6 mm in thickness. Due to their high sensitivity to air, the crystals were preserved and handled in an argon-filled glove box to minimize air exposure. Powder X-ray diffraction (XRD) was conducted using a Bruker D8 Advance diffractometer with Cu-Kα radiation. The orientation of the single crystals was confirmed using an Empyrean Malvern Panalytical X-ray diffractometer and Laue diffractometry with a PROTO Laue COS X-ray diffraction crystal orientation system. Compositional analysis and elemental mapping were performed using a field emission scanning electron microscope (FESEM TESCAN, MAGNA).

*Magnetic and electrical transport measurements.* The magnetic and ac susceptibility measurements were conducted utilizing a commercial Magnetic Property Measurement System (MPMS-3 SQUID Magnetometer, Quantum Design) with a maximum field capacity of 7 T. For in-plane magnetization measurements, the sample was affixed to a quartz paddle using a small quantity of Ge-varnish, whereas a brass holder was employed for out-of-plane measurements. The electrical transport measurements were executed using a commercial Physical Property Measurement System (PPMS, Quantum Design) with a maximum field capacity of 9 T. Resistivity measurements were performed via the standard four-probe technique, and Hall resistivity was assessed using a Hall bar configuration. To rectify data for contact misalignment, field symmetrisation was applied to magneto-resistivity, and field anti-symmetrisation was applied to the Hall resistivity data.

*Theoretical Calculations.* The first-principles calculations were performed by using the code of the Vienna ab initio simulation package (VASP) [1,2]. The exchange–correlation function was described by using the generalized gradient approximation (GGA), following the Perdew–

Burke–Ernzerhof (PBE) parametrization scheme [3]. The cut–off energy was set at 500 eV to expand the wavefunctions into a plane–wave basis. The convergence criteria for the total energy and ionic forces were set to $1\times10^{-5}$ eV and $-0.01$ eV Å$^{-1}$, respectively. The ferromagnetic order along the $c$ axis at the Gd sites were considered. To calculate the AHC, the ab initio density functional theory Bloch wavefunction was projected onto Maximally localized Wannier functions [4] with a diagonal position operator using VASP code. To obtain precise Wannier functions, we included the outermost $p$ orbitals for Bi, $d$ and $f$ orbitals for Gd, and $d$ orbitals for Ti to cover the full band overlap from the ab initio and Wannier functions. Based on the effective tight-binding model Hamiltonian, we performed the calculations for the AHC by the linear response Kubo formula approach in the clear limit as given by formula (1): [5]

$$\sigma_{xy} = \frac{e^2}{\hbar} \sum_n^{occ} \int \frac{d^3k}{(2\pi)^3} \Omega_{xy,n} \tag{1}$$

$$\Omega_{xy,n} = Im \sum_{m \neq n} \frac{\left\langle n \left| \frac{\partial H}{\partial k_x} \right| m \right\rangle \left\langle m \left| \frac{\partial H}{\partial k_y} \right| n \right\rangle - (x \leftrightarrow y)}{(E_{(n)} - E_{(m)})^2} \tag{2}$$

in formula (1) and formula (2), $\Omega_{xy,n}$ denoted the $xy$ component of the Berry curvature (BC) of the $n$th band, $n\rangle$ and $m\rangle$ are the eigenstates of $H$, and $E_{(n)}$ and $E_{(m)}$ are the corresponding eigenvalues. To realize integrations over the Brillouin zone, a $k$ mesh of $240 \times 240 \times 240$ points was used. The calculated electronic band structures of GdTi$_3$Bi$_4$ along high-symmetry paths clearly exhibit large Fermi pockets from both electrons and holes, resulting in a substantial density of states at the Fermi energy. The calculated magnetic moment is approximately 6.77 $\mu_B$/f.u., which aligns with the experimental findings.

## Structural characterization:

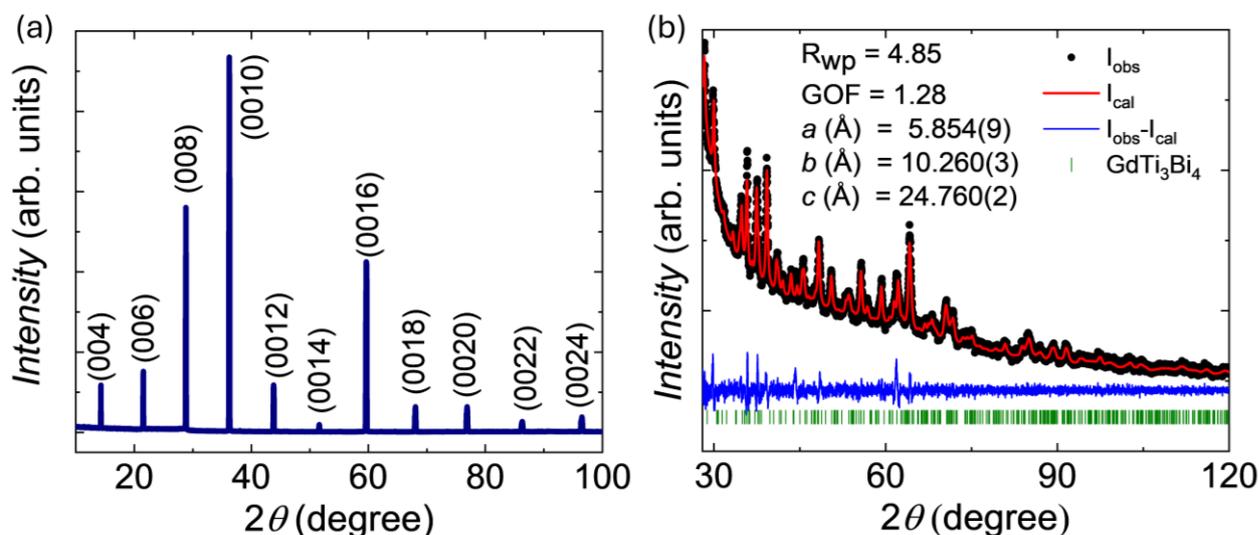

**Figure S1.** (a) XRD of GdTi$_3$Bi$_4$ single crystal revealing [00$l$] reflections. (b) The $hkl$ phase refinement of GdTi$_3$Bi$_4$ using TOPAS software. Observed intensity is shown by the black curve, the red curve indicates the calculated intensity, the difference between the observed and calculated intensity is shown by the blue line, and the green vertical bars represent the allowed Bragg reflections for GdTi$_3$Bi$_4$.

The XRD pattern of GdTi$_3$Bi$_4$ single crystal, shown in Figure S1 (a) exhibits sharp peaks exclusively of (00$l$) planes, indicating high crystalline quality and $c$-axis growth. Powder X-ray diffraction (PXRD) of GdTi$_3$Bi$_4$ was taken at room temperature using Cu K$\alpha$ X-rays as shown in Figure S1 (b). The GdTi$_3$Bi$_4$ single crystals are highly air-sensitive and are stored in an argon-filled glove box. Single crystals of GdTi$_3$Bi$_4$ were crushed and ground inside the glove box and placed in a XRD holder and carefully sealed with a mylar foil to protect it from oxidation. GdTi$_3$Bi$_4$ crystallizes in the orthorhombic crystal system with space group *Fmmm* (69). The lattice parameters obtained are $a$ = 5.854(9) Å, $b$ = 10.260(3) Å and $c$ = 24.760(2) Å. The presence of peaks in Figure S1 (b) only corresponding to GdTi$_3$Bi$_4$ *Fmmm* space group indicates high phase purity of sample. X-ray diffraction (XRD) refinement of crushed crystal powder confirms high phase purity of the sample and energy-dispersive X-ray spectroscopy (EDX) reveals a stoichiometric and homogeneous elemental distribution (see Figures S1 and S2).

# Compositional analysis:

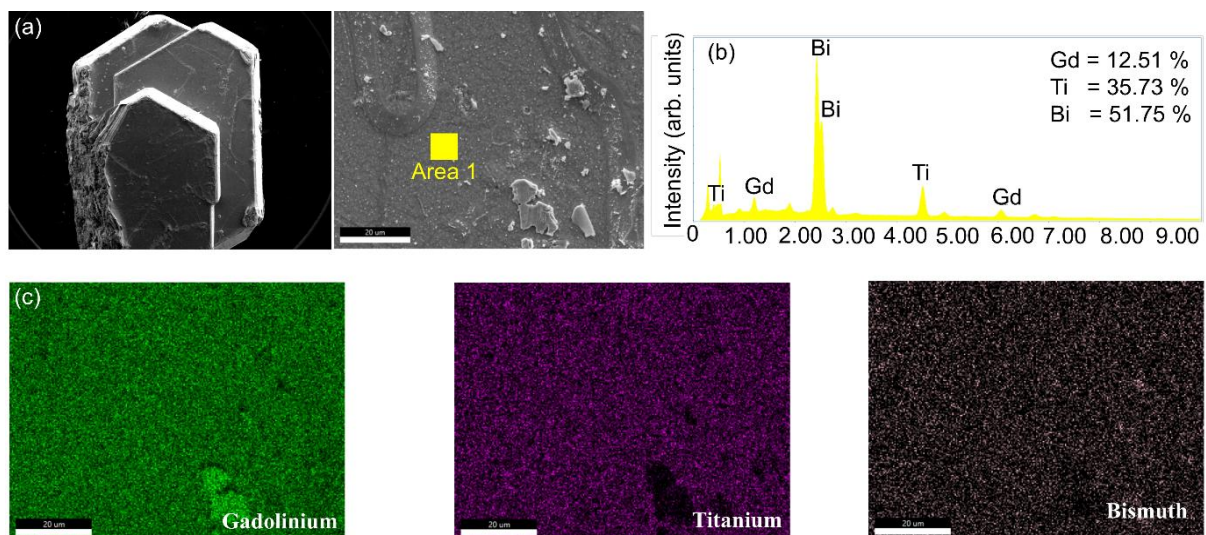

**Figure S2.** (a) SEM picture of GdTi$_3$Bi$_4$. (b) EDX spectra of GdTi$_3$Bi$_4$ with respective elemental atomic%. (c) Elemental mapping of Gd, Ti, and Bi.

The SEM image of analysis of GdTi$_3$Bi$_4$ as shown in Figure S2 (a) shows a layered growth of crystal stack planes. The EDX spectra taken at area 1 of crystal is shown in Figure S2 (b) with corresponding elemental atomic percentages. The average of EDX spectra taken at five different areas of sample gives average elemental proportion as GdTi$_{3.03}$Bi$_{4.04}$. The elemental mapping of Gd, Ti, and Bi as shown in Figure S2(c) suggests a homogeneous distribution of the constituent elements.

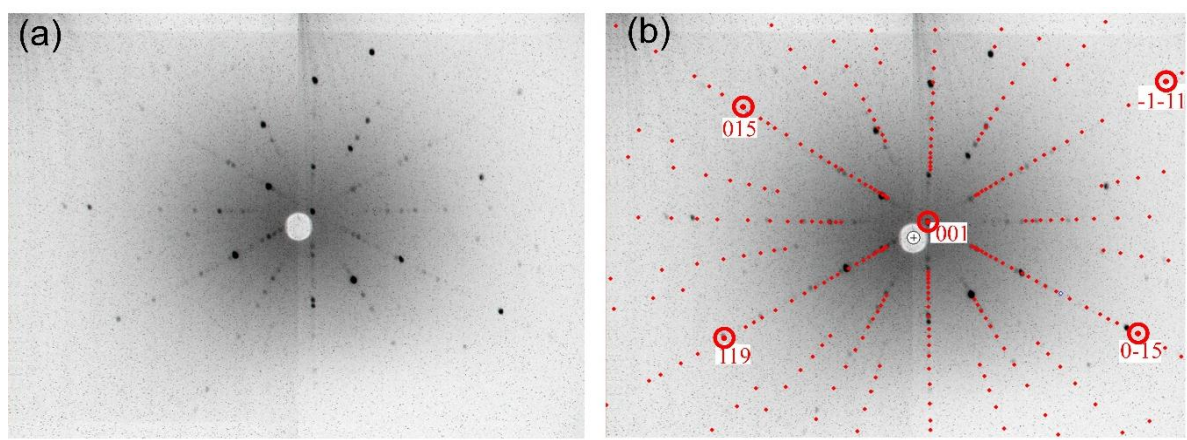

**Figure S3. Laue X-ray diffraction** (a)The experimental Laue diffraction pattern along [001]. (b) The obtained pattern is superposed with the simulated pattern along [001].

The obtained Laue diffraction pattern shown in Figure S3 (a) and the corresponding fitted pattern shown in Figure S3 (b) confirms the single crystalline nature of the sample without any twining.

## Magnetization measurements:

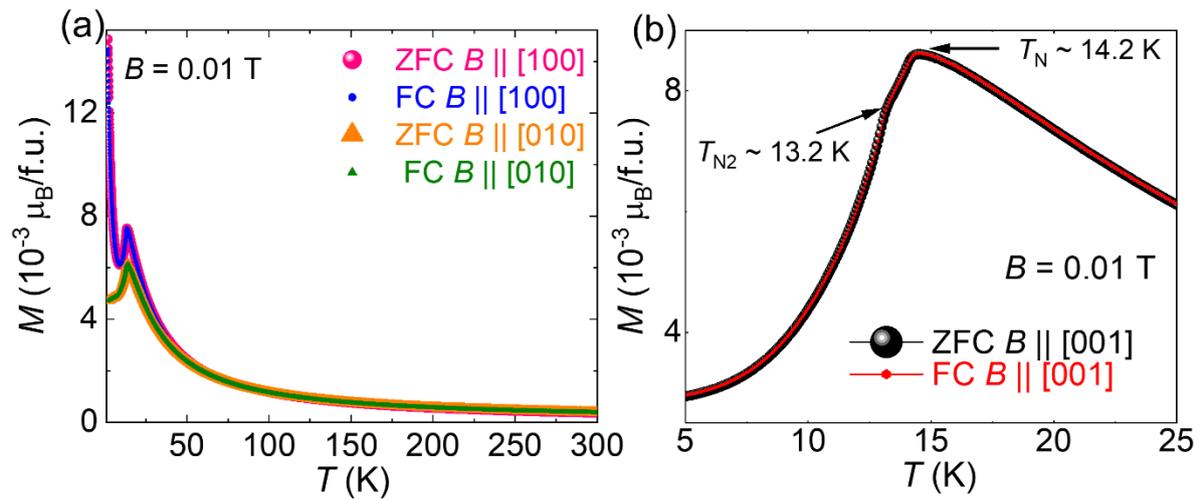

**Figure S4.** (a) Zero field-cooled (ZFC) and field-cooled (FC) dc magnetization $M(T)$ at 0.01 T magnetic field for $B$ parallel to [100] and [010] directions, respectively. (b) Dc magnetization $M(T)$ at 0.01 T for $B \parallel$ [001] showing $T_N$ =14.2 K and $T_{N2}$ =13.2 K.

Figure S4 (a) illustrates the temperature-dependent zero-field-cooled (ZFC) and field-cooled (FC) dc magnetization ($M(T)$) measured with the magnetic field applied along [100] and [010] crystallographic directions. Figure S4 (b) shows the $T_N$ and $T_{N2}$ for $B \parallel$ [001] as discussed in the manuscript.

# Curie-Weiss (CW) fit of dc inverse susceptibility along [100] and [010] crystallographic directions:

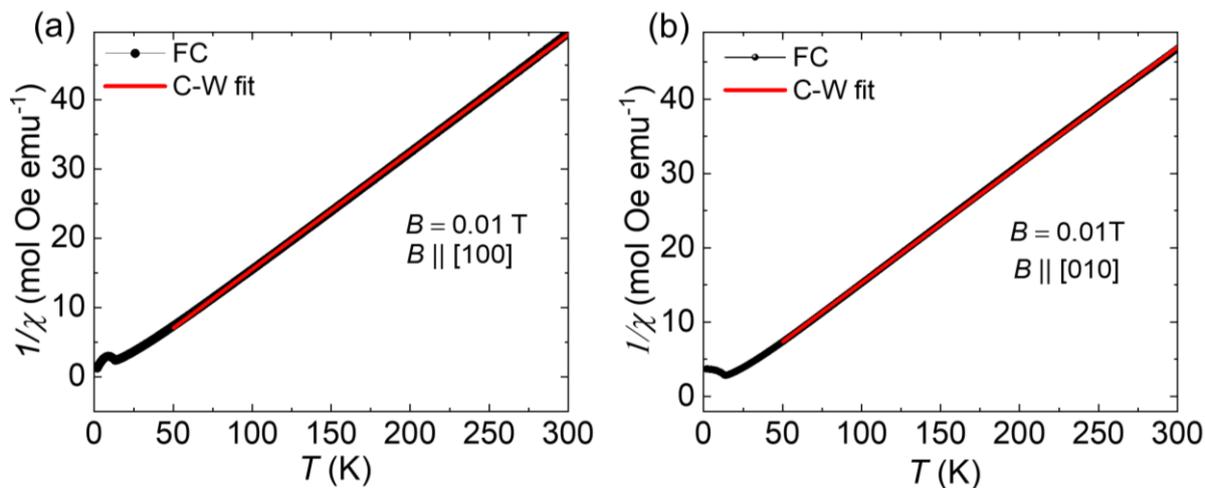

**Figure S5.** The CW fit of dc inverse susceptibility ($1/\chi$) vs $T$ for (a) $B \parallel a$ and (b) $B \parallel b$.

The CW fit has been performed for $1/\chi(T)$ along the crystallographic axes [100] and [010] as shown in Figure S5 (a) and S5 (b) respectively. The fit yields $\theta_p \sim 8.69$ K and 3.19 K for $B \parallel$ [100] and $B \parallel$ [010], respectively. The calculated $\mu_{eff} \sim 6.89$ (for $B \parallel$ [100]) and 7.10 $\mu_B$ (for $B \parallel$ [010]) are in good agreement with the expected theoretical value $\left(gJ(J+1)_{\mu B}\right)$ of 7.93 $\mu_B$/Gd$^{3+}$.

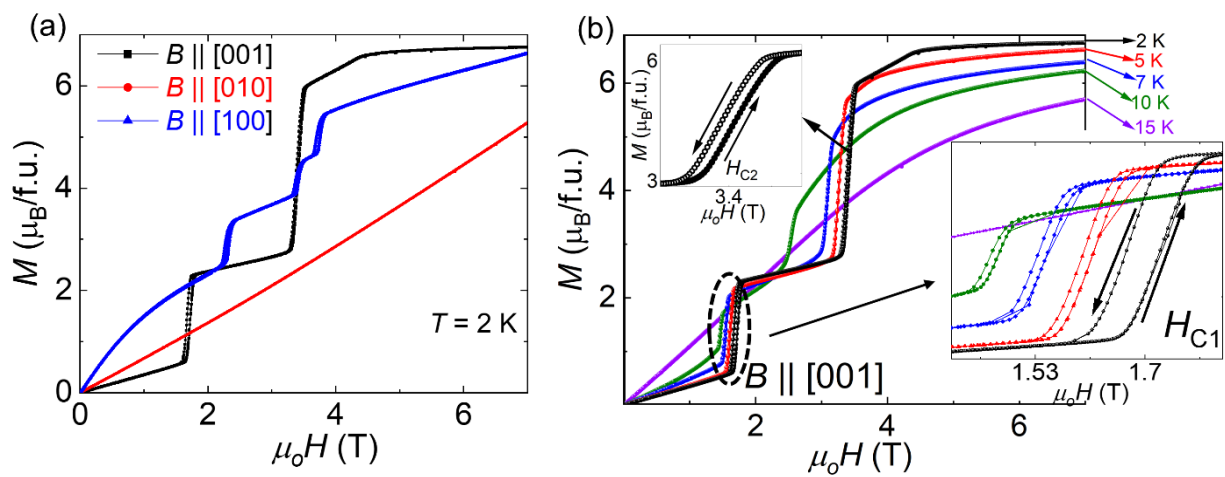

**Figure S6.** (a) Field-dependent magnetization $M(H)$ at $T = 2$ K for $B$ parallel to [100], [010], and [001] directions, respectively. (b) Isothermal magnetization $M(H)$ for $B \parallel$ [001] at different temperatures, the (lower and upper) inset shows zoomed in view of the hysteresis at $H_{C1} \sim 1.7$ T at different temperatures, $H_{C2} \sim 3.4$ T at 2 K respectively.

The corresponding field-dependent magnetization measured along various crystallographic axes at $T = 2$ K are shown in Figure S6 (a). Multiple metamagnetic transitions are observed when the magnetic field is applied along [001] and [100], whereas no such transitions are observed along [010]. Figure S6 (b) illustrates the isothermal magnetization $M(H)$ for $B \parallel$ [001] at different temperatures, the zoomed-in view of $H_{C1} \sim 1.7$ T at different temperatures, $H_{C2} \sim 3.4$ T at 2 K is shown in the inset.

## Theoretical calculations:

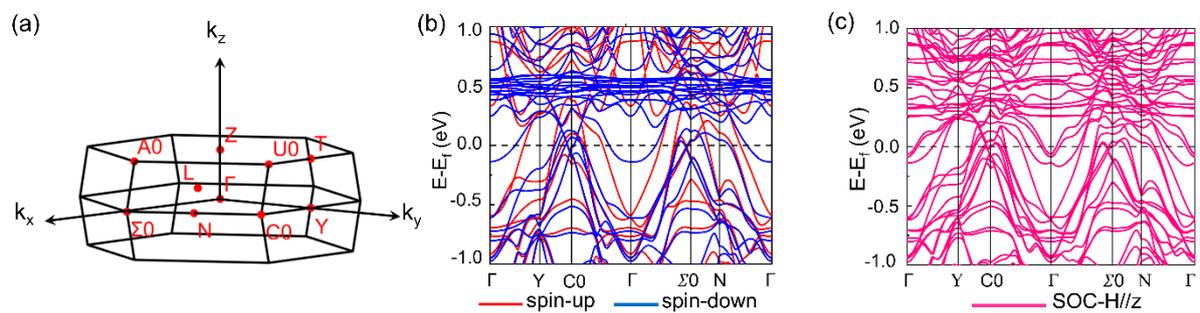

**Figure S7.** (a) Three-dimensional Brillouin zone of the primitive cell. (b) Energy dispersion of GdTi$_3$Bi$_4$ with FM state and without considering spin-orbital coupling (SOC), where red and blue lines are for spin-up and spin-down channels, respectively. (c) Energy dispersion with magnetization along [001] and incorporating the SOC.

GdTi$_3$Bi$_4$ crystallizes in an orthorhombic lattice with space group *Fmmm* (No. 69), and its three-dimensional Brillouin zone for the primitive cell is depicted in Figure S7 (a). The zig-zag chains of Gd offers distinct magnetic exchange pathways and the kagome layers of Ti, through their associated electronic band structures, together influence the material's magnetic and transport properties. The first-principles based band structure calculations indicate presence of multiple linear and flat bands near Fermi level as illustrated in Figure S7(b) and S7 (c), (without and with spin-orbit coupling, respectively).

**Derivative of magnetization with respect to applied magnetic field (d*M*/d*H*) at different temperatures:**

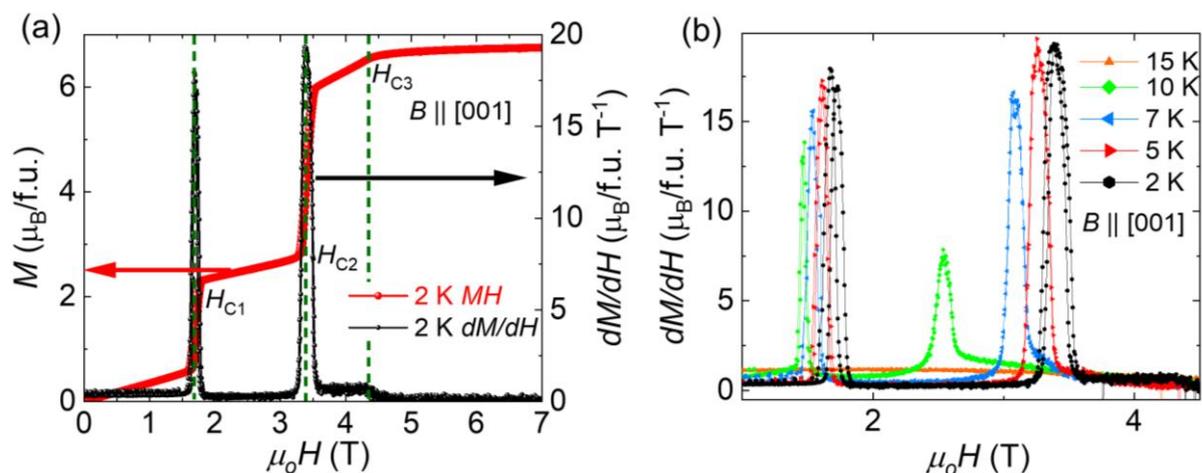

**Figure S8.** (a) *M*(*H*) on the left scale and *dM/dH* on the right scale for *B* ∥ [001] at *T* = 2 K three magnetic transitions are clearly observed. (b) *dM/dH* for *B* ∥ [001] at different temperatures.

Figure S8 (a) shows the *M*(*H*) and *dM/dH* at 2 K along [001], we can clearly observe the metamagnetic transitions at $H_{C1}$ = 1.7 T, $H_{C2}$ = 3.4 T and $H_{C3}$ = 3.95 T. *dM/dH* at different temperatures along [001] are shown in Figure S8 (b). The critical fields corresponding to the metamagnetic transitions decreases with increase in temperature as depicted in Figure S8 (b).

## Temperature and field-dependent magnetization along [100] and [010] crystallographic directions:

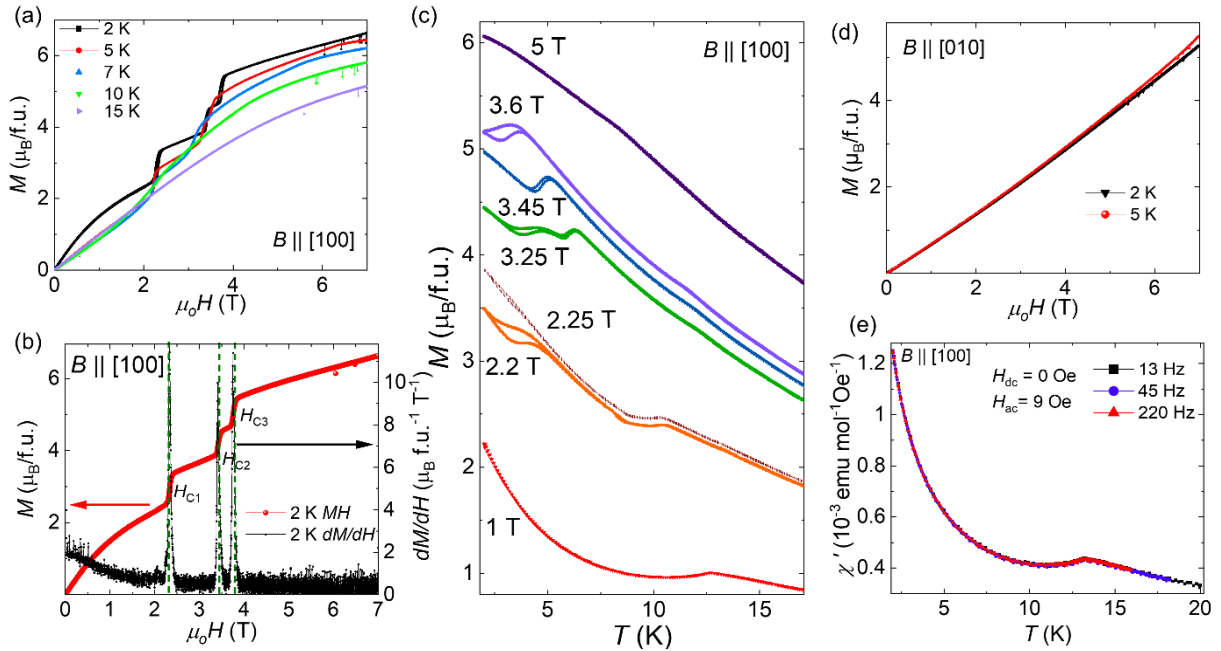

**Figure S9.** (a) $M(H)$ at different temperatures for $B \parallel [100]$. (b) $M(H)$ on the left scale and $dM/dH$ on the right scale for $B \parallel [100]$ at 2 K three metamagnetic transitions are clearly observed. (c) Temperature dependent FCC and FCW curves at different magnetic fields for $B \parallel [100]$. (d) $M(H)$ at 2 K and 5 K for $B \parallel [010]$. (e) In-phase component of ac susceptibility measured along [100] at different frequencies and $H_{ac}$ = 9 Oe.

Figure S9 (a) shows isothermal magnetization ($M(H)$) along [100] at different temperatures. We observe three metamagnetic transitions when $B \parallel [100]$, $H_{C1}$ = 2.35 T, $H_{C2}$ = 3.4 T and $H_{C3}$ = 3.79 T as shown in Figure S9 (b). However, the magnetization along [100] does not saturate up to an applied magnetic field of 7 T. Figure S9 (c) shows the temperature-dependent FCC and FCW curves at various applied magnetic fields along [100] in the vicinity of the three metamagnetic transitions. Around the critical fields $H_{C1} \sim 2.35$ T and $H_{C2} \sim 3.4$ T, thermal hysteresis is observed between the field-cooled cooling (FCC) and field-cooled warming (FCW) magnetization curves, indicating a first order phase transition. We could not observe any metamagnetic transitions when $B \parallel [010]$ as shown in Figure S9 (d). Along [010], the magnetization increases linearly and does not saturate up to an applied magnetic field of 7 T. Figure S9 (e) shows the temperature-dependent ac susceptibility $\chi'(T)$ for $B \parallel [100]$ and $H_{dc}$ = 0 T, the peak at $T_N \sim 14.2$ K in real part of $\chi'(T)$ does not show any shift in temperature upon varying the excitation frequency.

**Temperature dependent ac susceptibility measurements along [001]:**

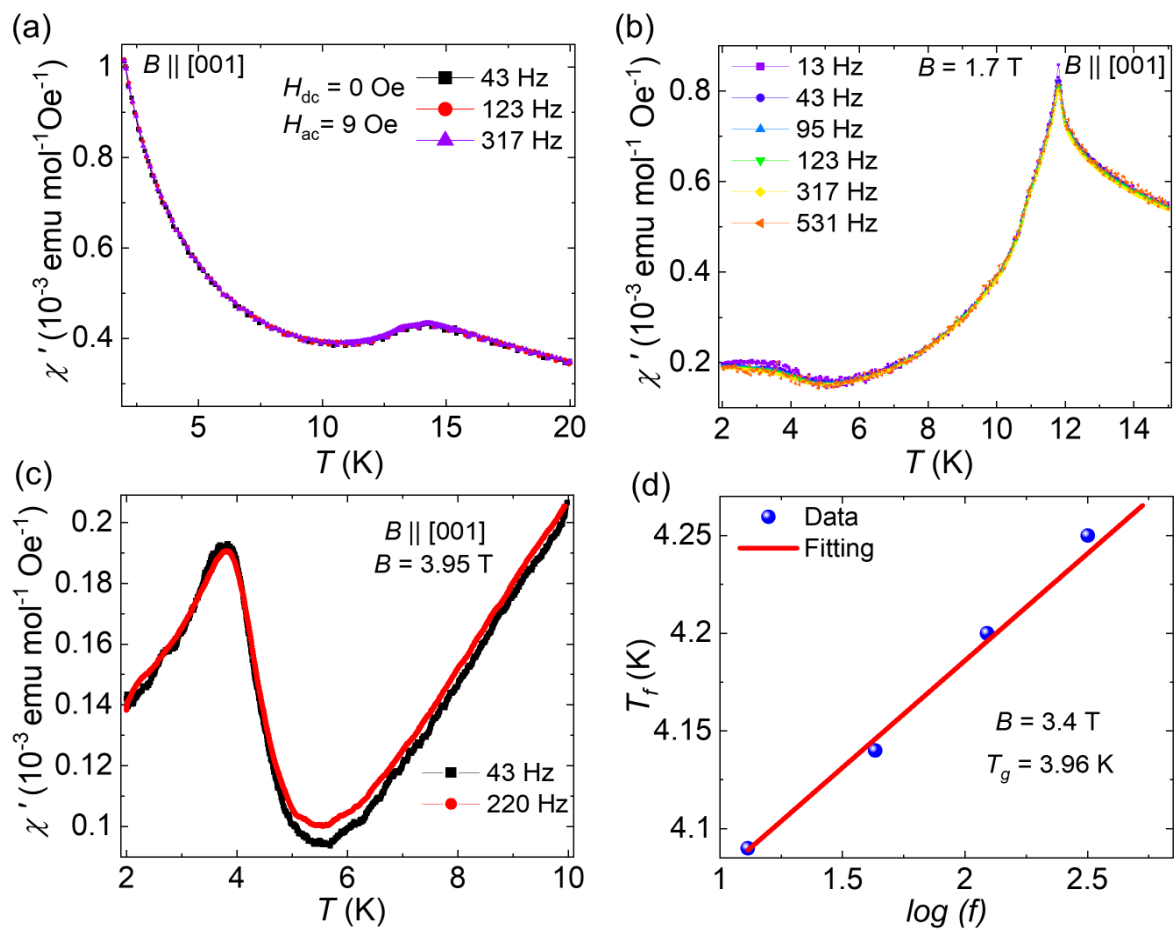

**Figure S10.** In-phase component of temperature dependent ac susceptibility $\chi'(T)$ measured along [001] at different frequencies, $H_{ac} = 9$ Oe, (a) $H_{dc} = 0$ T, (b) and (c) 1.7 T and 3.95 T, respectively. (d) Linear fit of $T_f$ vs log $(f)$ to estimate the value of $T_g$ for $H_{dc} = 3.4$ T.

Figure S10 (a) shows the temperature-dependent ac susceptibility $\chi'(T)$ at various excitation frequencies $f$ of ac amplitude ($H_{ac} = 9$ Oe). For $B \parallel c$ and $H_{dc} = 0$ T, the peak in real part of $\chi'(T)$ does not exhibit any frequency dependence. This rules out the possibility of the presence of any type of typical glassy transition. Figure S10 (b) and S10 (c) shows real component of ac susceptibility measured along [001] at various frequencies, $H_{ac} = 9$ Oe, $H_{dc} = 1.7$ T, $\chi'(T)$ shows two maxima and no change in peak position with variation in $f$. This suggests that the lower field-induced first-order phase transition does not exhibit any frequency dependent spin dynamics. Similarly, at $H_{dc} = 3.95$ T as shown in Figure S10 (c) $\chi'(T)$ do not show any shift in peak upon varying the excitation frequency, indicating absence of glassy transition or magnetic frustrations. It is important to note that with increasing dc field $T_N$ shifts to a lower temperature, so we observe a peak at 3.5 K when $H_{dc} = 3.95$ T is applied (further explained in the main text). In Figure S10 (d), we illustrate the frequency dependent peak shift for 3.4 T applied dc field, where the first order phase transition is observed in Figure 2 (a) of the main manuscript. Here $T_f$ varies linearly with log ($f$) and after extrapolation of the linear fit in the curve, we can estimate the value of $T_g \sim 3.96$ K.

**Electrical transport measurements: Schematic of mechanism involved in the magnetoresistance (MR) measured when magnetic field was applied along [001] and current was applied along [100]:**

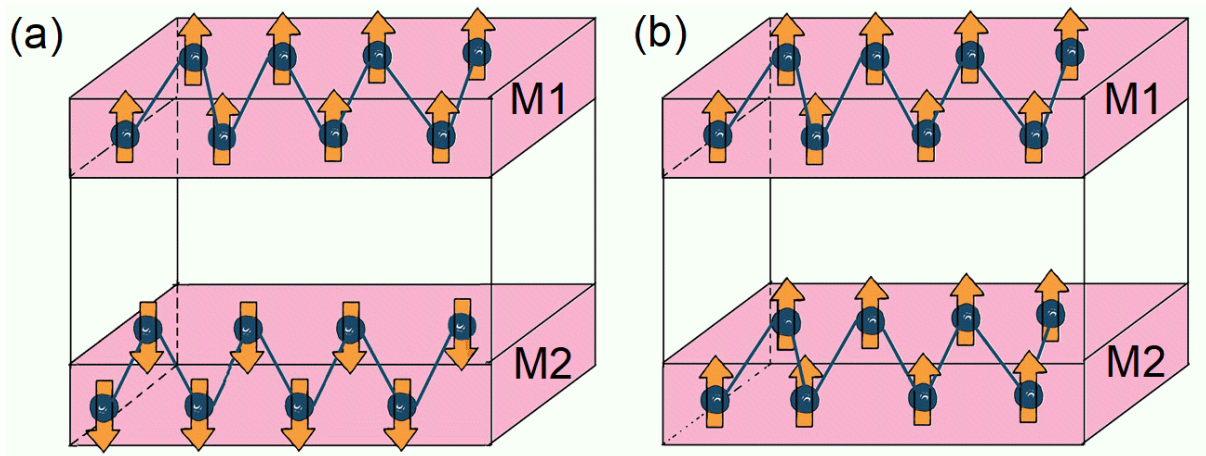

**Figure S11.** Schematic of adjacent magnetic layers forming (a) high resistance state [6] and (b) low resistance state.

At $H_{C1}$ the MR% increases by 26.6% while at $H_{C2}$ the MR% decreases by 27.7%. At $H_{C1}$, the onset of the first magnetization plateau, leads to the development of high-resistance state [6] i.e. the adjacent magnetic layers (M1 M2) have opposite moment orientation as shown in schematic Figure S11 (a), At $H_{C2}$, the onset of second plateau leads to a low resistance state with moments of both magnetic layers aligned in same direction, as shown in schematic Figure S11 (b). This gives rise to jumps in MR commonly observed in GMR multilayers [7,8].

## Detailed procedure for extracting $\rho_{yx}^o$, $\rho_{yx}^A$, $\rho_{yx}^T$ :

We have quantitatively described the subtraction of metamagnetic jumps in $\rho_{yx}$ via linear extrapolation of the ordinary Hall data immediately adjacent to these jumps, with the full protocol outlined step-by-step. Representative raw and processed Hall resistivity curves are now shown for the 7 K and 15 K datasets in Figure S12, illustrating the extraction of the ordinary, anomalous, and topological components.

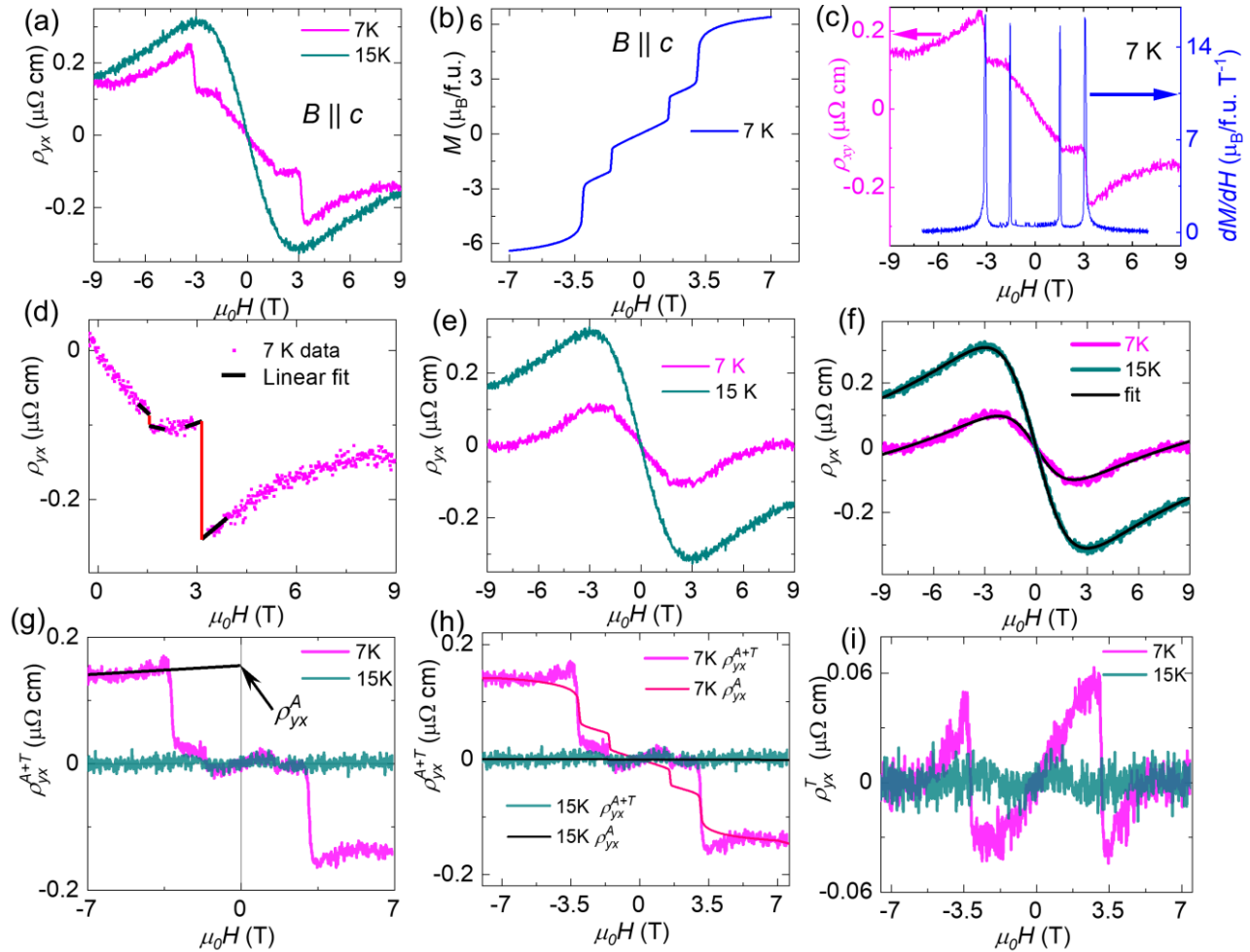

Figure S12. Magnetic field dependence of (a) $\rho_{yx}$ at 7 K and 15 K, and (b) Magnetization M at 7 K ($B \| c$). (c) $\rho_{yx}$ and derivative $dM/dH$ at 7 K. (d) Magnetic field dependence of $\rho_{yx}$, black lines show linear fits to the data near $H_{C1}$ and $H_{C2}$; the red lines guide the eye in showing the jumps in $\rho_{yx}$ caused by magnetization jumps. After subtracting these jumps (shift indicated by red lines in (d)), we obtain (e) the ordinary Hall resistivity ($\rho_{yx}^o$) at 7 K, alongside paramagnetic-state data at 15 K ($T > T_N$), dominated by two band behaviour. (f) Nonlinear $\rho_{yx}^o$ fitted using a two-band model. Subtracting $\rho_{yx}^o$ from $\rho_{yx} = \rho_{yx}^o + \rho_{yx}^A + \rho_{yx}^T$ yields (g) $\rho_{yx}^{A+T}$, the high field data is linear fitted extrapolated to zero field to obtain anomalous Hall resistivity $\rho_{yx}^A$ part (h).

$\rho_{yx}^{A+T}$ is further plotted with the calculated anomalous Hall part using magnetization. (i) Field dependent additional Hall resistivity ($\rho_{yx}^{T}$) at different temperatures.

According to Matthiessen's rule the total Hall resistivity is given by:

$$\rho_{yx} = \rho_{yx}^{o} + \rho_{yx}^{A} + \rho_{yx}^{T} \quad (1)$$

Where $\rho_{yx}^{o}$ is ordinary Hall resistivity due to Lorentz force, $\rho_{yx}^{A}$ is anomalous Hall resistivity from magnetization and $\rho_{yx}^{T}$ represents additional Hall resistivity arising from the topological Hall contributions from trivial or non-trivial spin textures etc. As illustrated in Figure S12(a), the magnetic field dependence of Hall resistivity ($\rho_{yx}$) below $T_N$ =14.2 K, specifically at 7 K, exhibits nonlinearity and abrupt jumps induced by magnetization in the antiferromagnetic state. Conversely, above $T_N$ at 15 K, within the paramagnetic regime, the Hall resistivity displays a continuous curve without jumps, yet it demonstrates nonlinearity with both positive and negative slopes, suggesting the presence of multiple charge carriers or bands. The ordinary Hall resistivity $\rho_{yx}^{o}$, is anticipated to manifest as continuous curves rather than discontinuous jumps. The observed jumps are attributed to the superposition of anomalous Hall resistivity $\rho_{yx}^{A}$, which causes significant jumps in $\rho_{yx}$ due to changes in magnetization. The magnetization at 7 K is depicted in Figure S12(b), showing jumps at high-field and the 1/3 plateaus. These jumps in $\rho_{yx}$ precisely correspond to the jumps in $M(H)$, as demonstrated in the comparative plot of Hall resistivity and the field derivative of magnetization shown in Figure S12(c). By analyzing the Hall data from the region where the derivative of magnetization is nearly constant, as seen in Figure S12(c), and excluding data points where magnetization peaks due to substantial changes at $H_{C1}$ and $H_{C2}$, the results are plotted in Figure S12(d).

Now, to account for the discontinuities induced by metamagnetic transitions (MMT), the $\rho_{yx}$ signal was adjusted by subtracting a linear fit of the data dominated by ordinary behavior, specifically in the region characterized by a nearly constant magnetization derivative around $H_{C1}$ and $H_{C2}$, as indicated by the black linear fit. The discontinuities in Hall resistivity, attributed to magnetization jumps, are represented by a red straight line. These hurdles were removed by subtracting the red jumps, which correspond to the intercepts of the linear fits at $H_{C1}$ and $H_{C2}$. Following this subtraction, as depicted in Figure S12(e), the extracted ordinary component of the 7 K Hall resistivity is presented alongside the ordinary raw data at 15 K in the paramagnetic state. This data exhibits similar nonlinear Hall behavior, indicating that the subtraction accurately reflects the ordinary Hall characteristic behavior across the field range.

The data for all temperatures is presented in manuscript Figure 3(d). Now the 7K and 15K data was then fitted with the two-band model given as

$$\rho_{yx}^o = \frac{H}{e} \frac{(n_h\mu_h^2 - n_e\mu_e^2) + (n_h - n_e)\mu_h^2\mu_e^2 H^2}{(n_h\mu_h^2 + n_e\mu_e^2)^2 + (n_h - n_e)^2\mu_h^2\mu_e^2 H^2} \qquad (2)$$

The fitting is illustrated in Figure S12(f) and is reasonably good. Further this ordinary contribution $\rho_{yx}^o$ obtained from fitting is subtracted from $\rho_{yx} = \rho_{yx}^o + \rho_{yx}^A + \rho_{yx}^T$ and we obtained $\rho_{yx}^A + \rho_{yx}^T$ or $\rho_{yx}^{A+T}$ as shown in Figure S12 (g). At high fields the contributions from additional Hall resistivity ($\rho_{yx}^T$) arising from spin textures saturates. Therefore, the high field data is linear fitted and extrapolated to zero field to obtain the $\rho_{yx}^A$ at 7K as shown in Figure S12 (g). The anomalous Hall resistivity is given by $\rho_{yx}^A = 4\pi R_S M$. The intercept $\rho_{yx}^{A+T}$ in the high field regime, where $\rho_{yx}^T$ vanish and magnetization saturates, yields $4\pi R_s$ [1]. Using experimental magnetization, the anomalous Hall resistivity ($4\pi R_s M$) is calculated and depicted in Figure S12 (g). Subtracting this anomalous part $\rho_{yx}^A$ governed by magnetization from the $\rho_{yx}^{A+T}$ we obtained $\rho_{yx}^T$ At 7K and 15 K as shown in Figure S12 (i) .

## Carrier concentration and mobility for GdTi$_3$Bi$_4$:

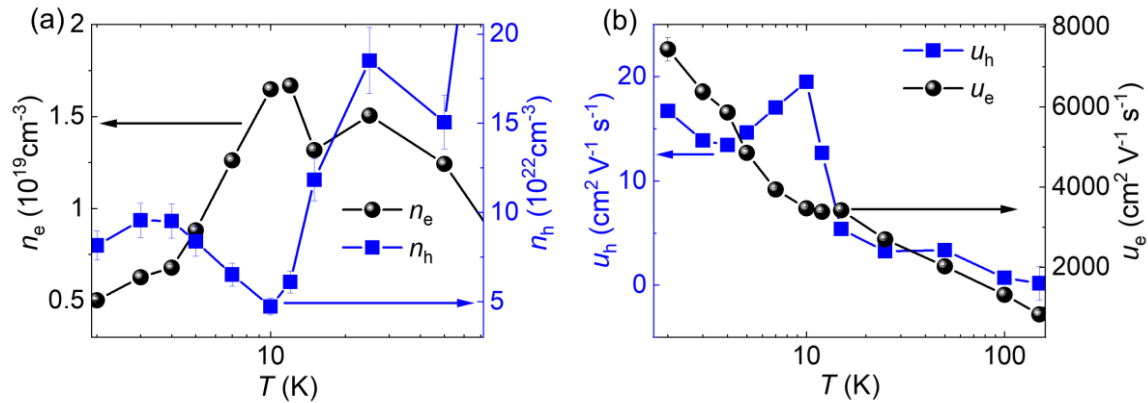

**Figure S13.** Temperature dependence of charge (a) carrier density and (b) carrier mobility.

The temperature dependence of charge carrier density and charge carrier mobility is shown in Figure S13 (a) and S13 (b), respectively. The field dependent Hall resistivity exhibits positive and negative slope due to contributions from both electrons and holes. The carrier density of electron band is relatively lower, and mobility of electrons is very high signifying small electron pockets at the Fermi level. On the contrary the high carrier density and low mobility of holes bands signify relatively large area of hole pockets at $E_F$. The carrier density increases while the mobility decreases with temperature.

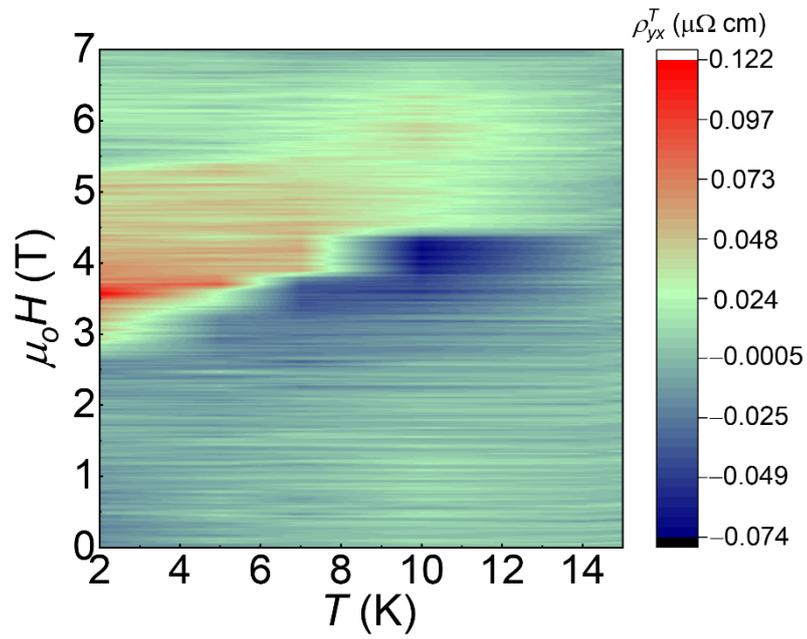

**Figure S14.** Contour plot of additional Hall resistivity arising from spin textures ($\rho_{yx}^{T}$) as a function of temperature and magnetic field.

**Analysis of anomalous Hall resistivity:**

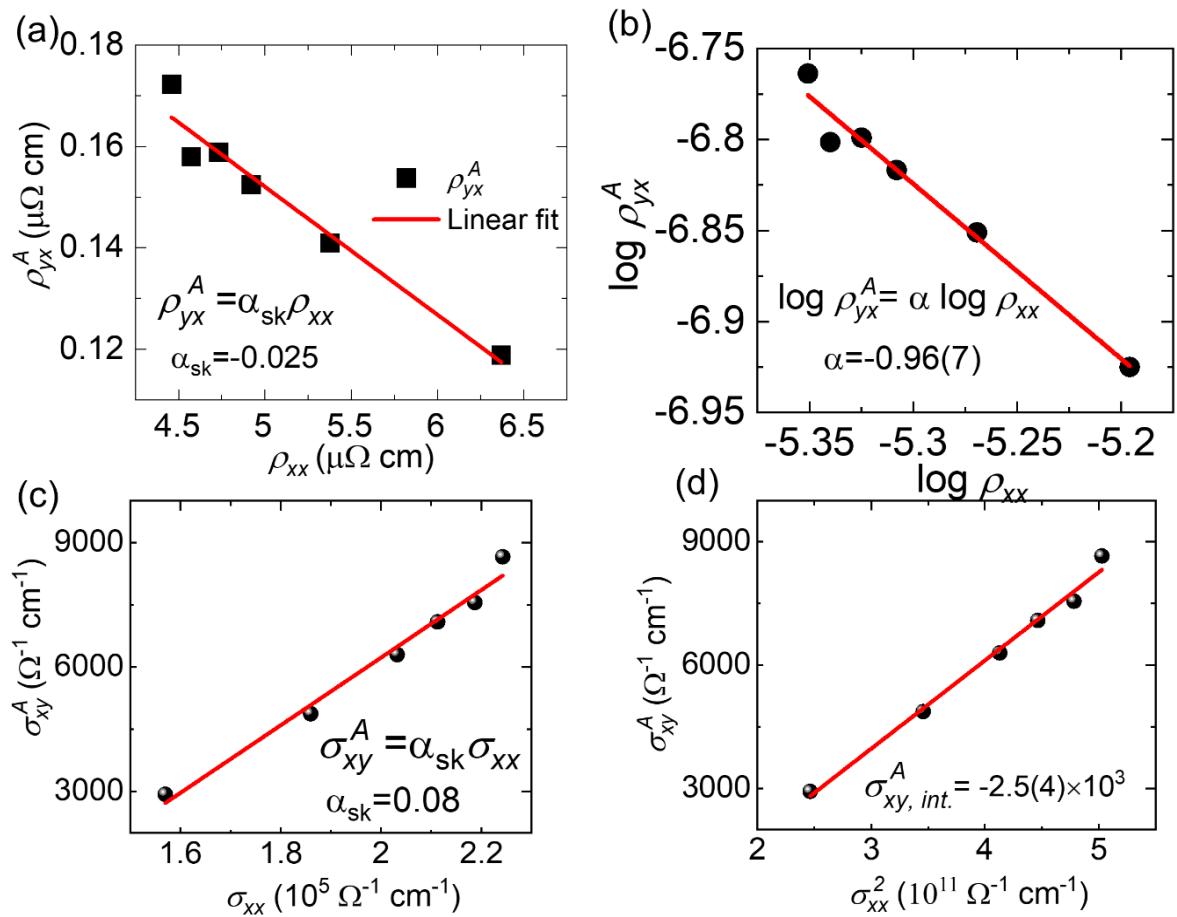

**Figure S15.** (a) Plot of anomalous Hall resistivity versus longitudinal resistivity with linear fit, (b) Logarithmic plot, (c) plot anomalous Hall conductivity (AHC) versus longitudinal conductivity, and (d) TYJ model [9,10] for conductivity.

The conclusions of dominance of skew scattering are further verified by other simpler models. As shown in Figure S15 (a) a linear relationship between $\rho_{yx}^A$ and $\rho_{xx}$ suggests dominance of skew scattering, from log plot as shown in Figure S15 (b) the slope $\alpha \sim 1$ suggests dominance of skew scattering and the same conclusion follows from linear dependence of conductivity shown in Figure S15 (c) . The slope $\alpha$ is the skew scattering constant and large slope $\alpha = 0.08$ signifies giant skew scattering. The TYJ model for conductivity shown in Figure S15 (d) is in good agreement with the TYJ plot of resistivity shown in the manuscript Figure 3h [11,12].

## Band calculation with different value of the Coulomb interaction U:

We investigated the influence of the Coulomb interaction on the electronic band structure and AHC of Gd, which contains *f*-electrons. The Coulomb interaction parameter was varied from 2 to 6 eV in increments of 0.5 eV. The calculated electronic band structures and AHC, both with and without the Coulomb interaction, are presented in Figure S16 (a-t). As shown in Figure S16 (a & b), in the absence of the Coulomb interaction, flat bands are situated approximately 0.5 eV above the Fermi level, and a corresponding peak in the AHC appears at the same energy. When the Coulomb interaction is included and progressively strengthened, these flat bands gradually shift away from the Fermi level. Consequently, the AHC peak also shifts and eventually converges to a value around 230 $\Omega^{-1}$ cm$^{-1}$ as shown in Figure S17.

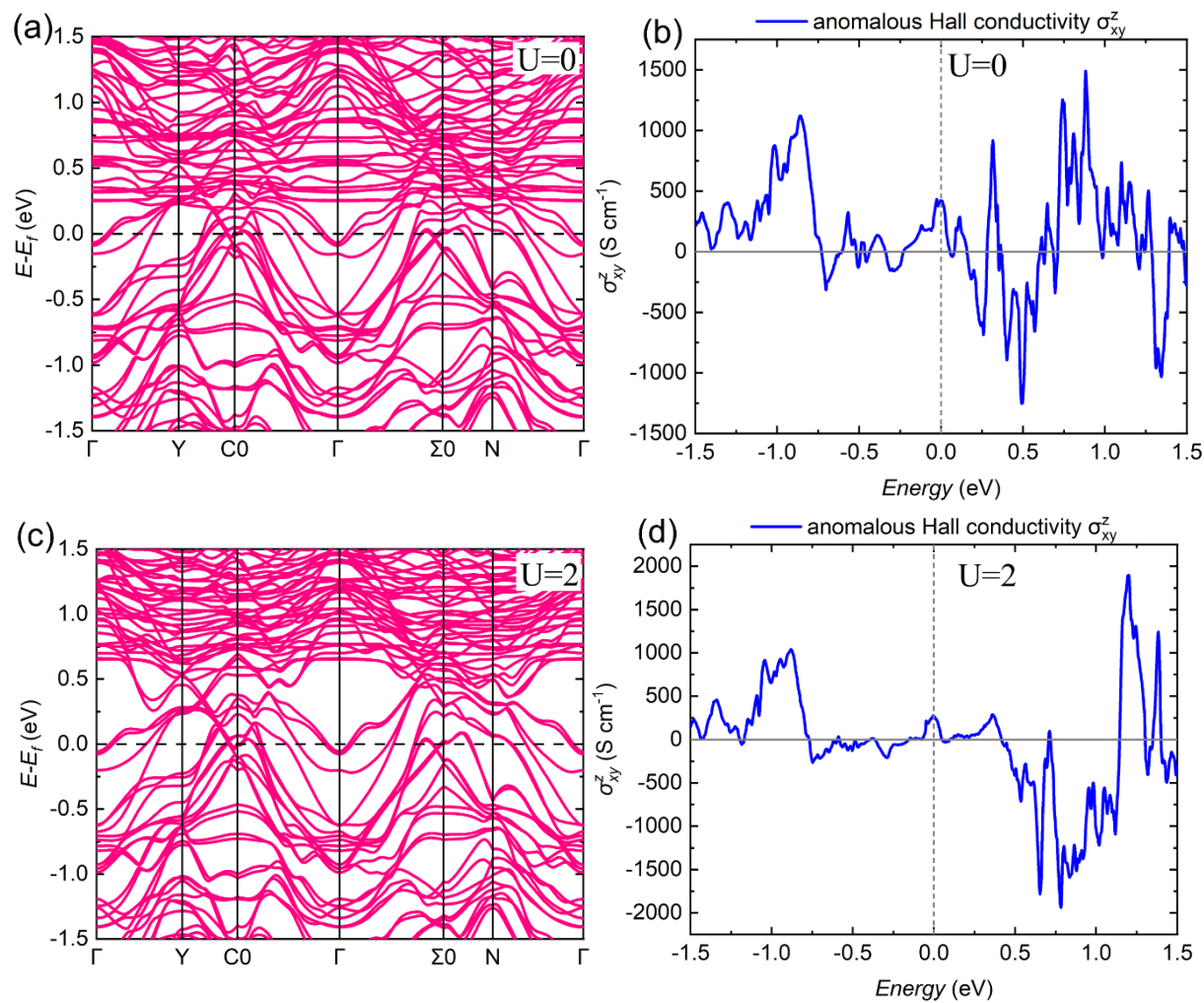

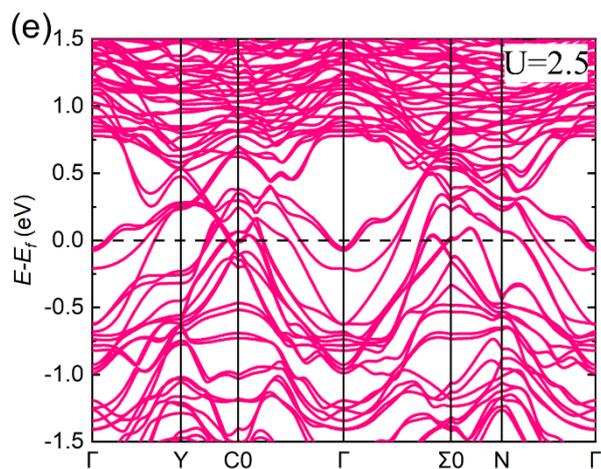

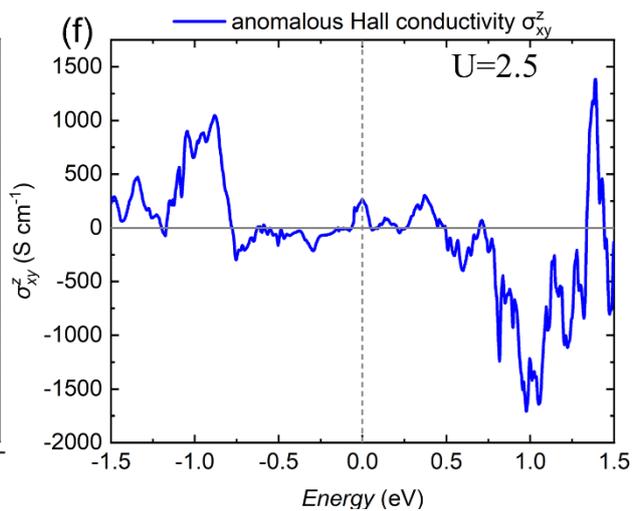

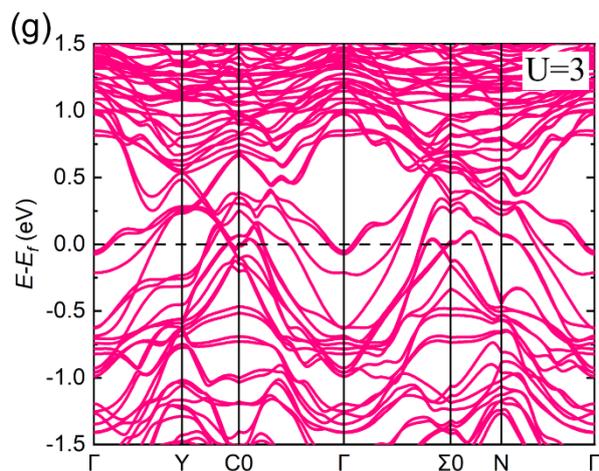

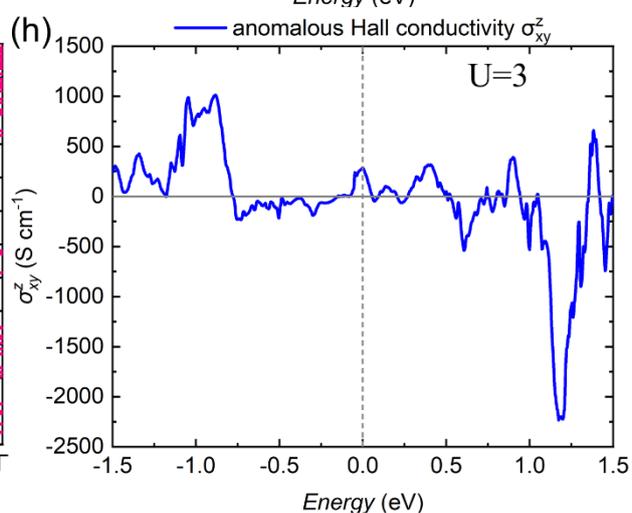

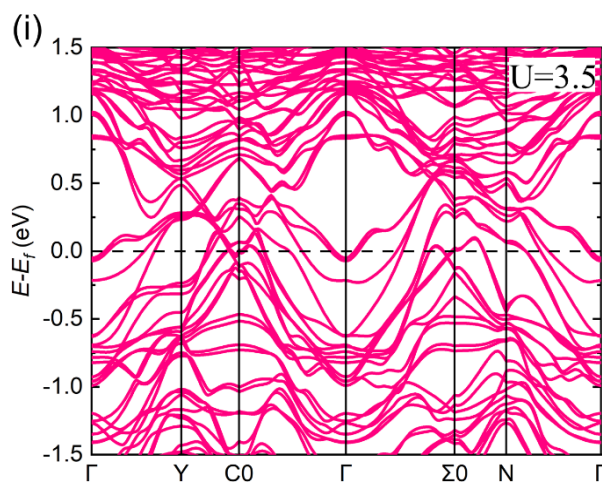

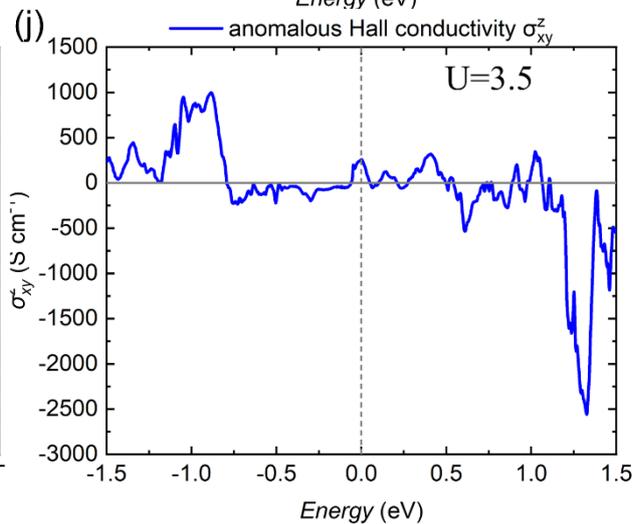

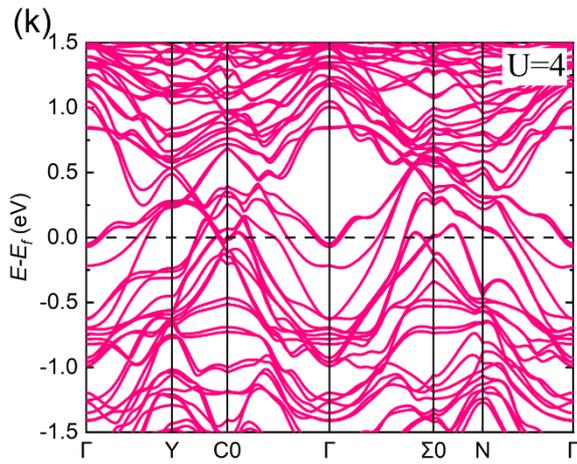

(k)

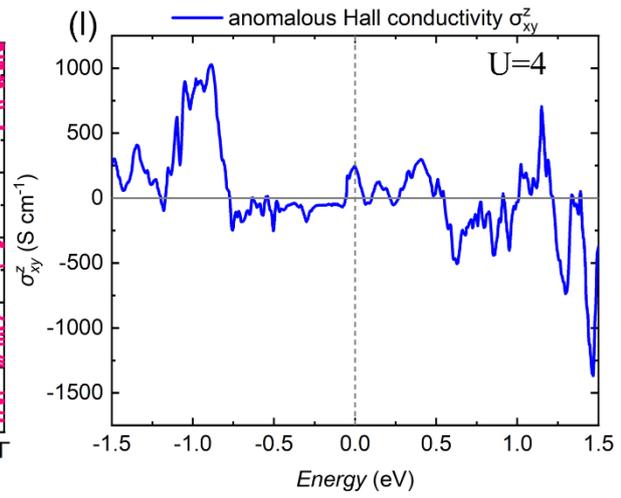

(l)

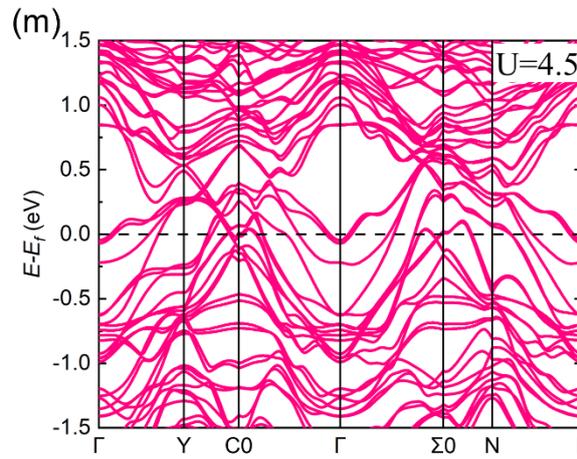

(m)

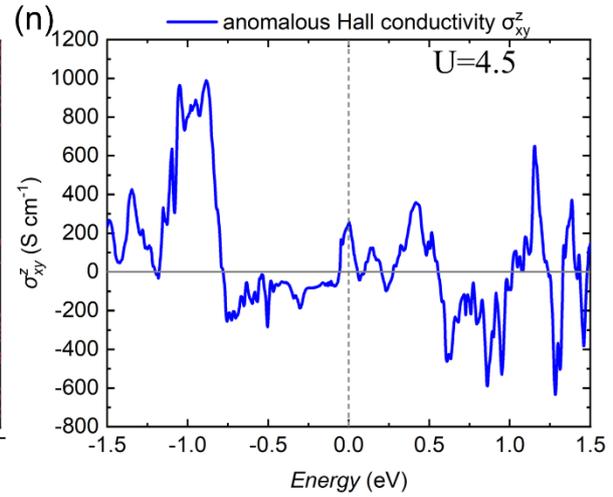

(n)

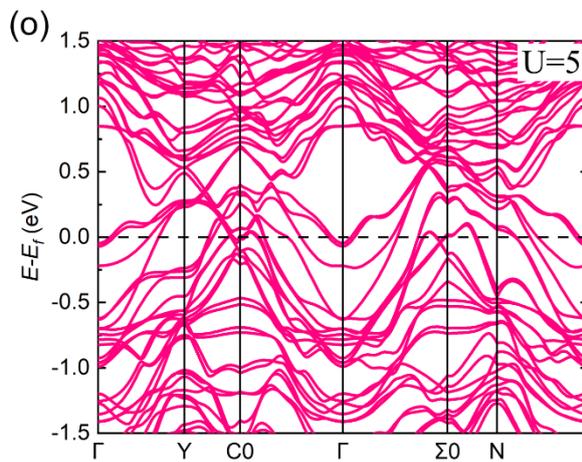

(o)

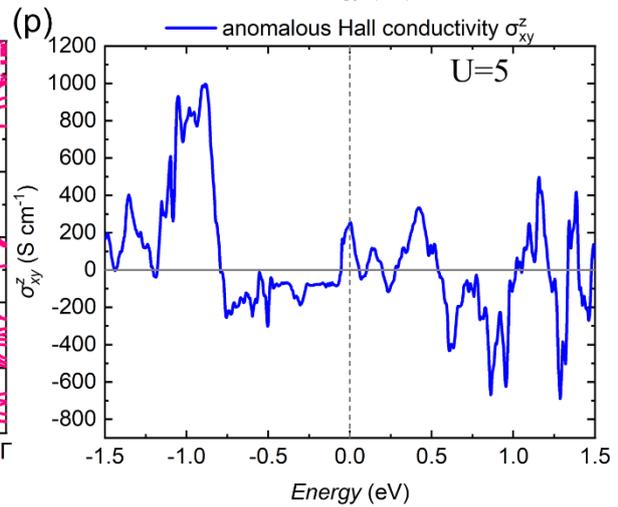

(p)

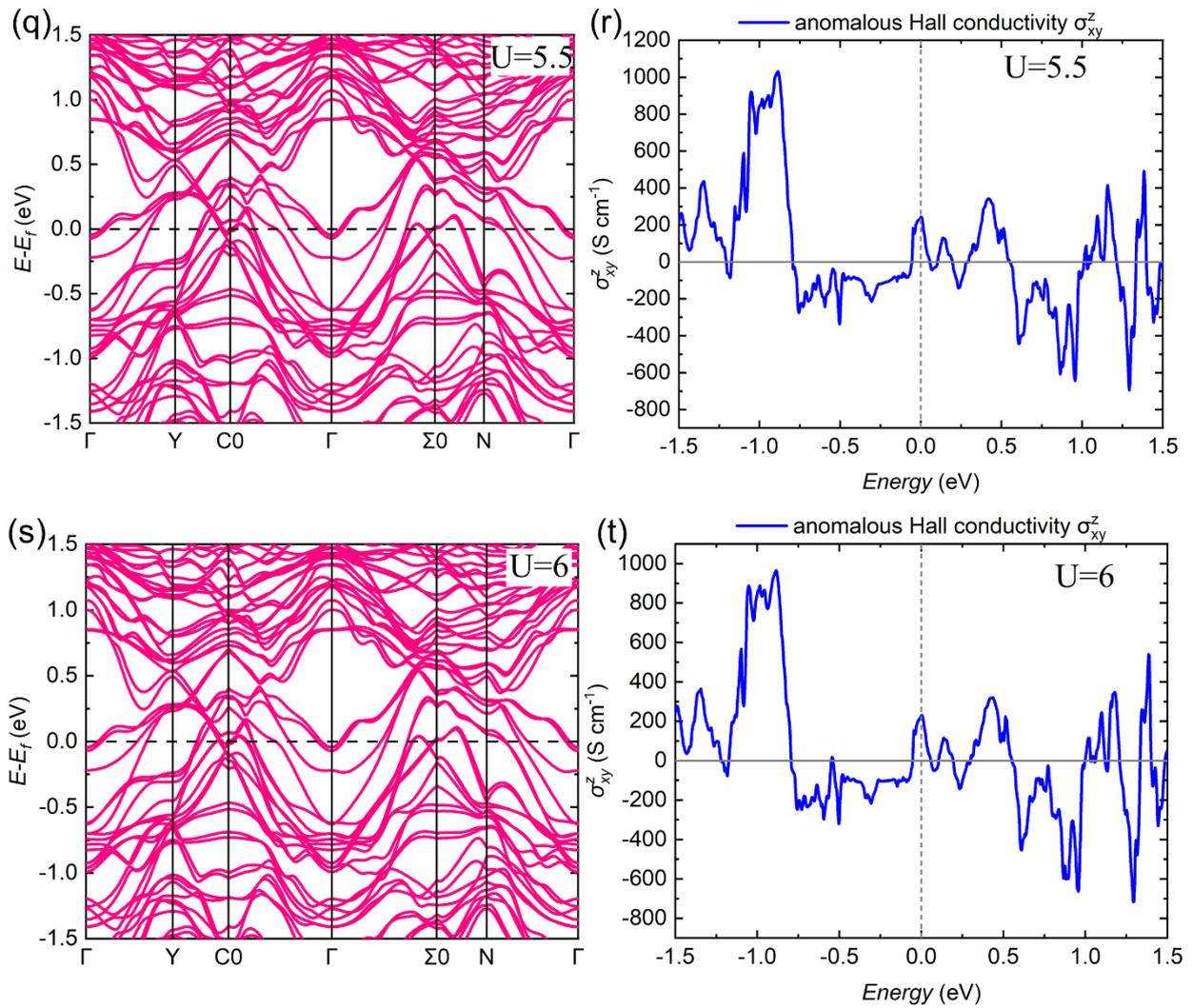

**Figure S16.** The calculated electronic band structures and AHC both with and without the Coulomb interaction, (a & b) U = 0, (c & d) U= 2, (e & f) U= 2.5, (g & h) U= 3, (i & j) U= 3.5, (k & l) U= 4, (m & n) U= 4.5, (o & p) U= 5, (q & r) U= 5.5, (s & t) U=6.

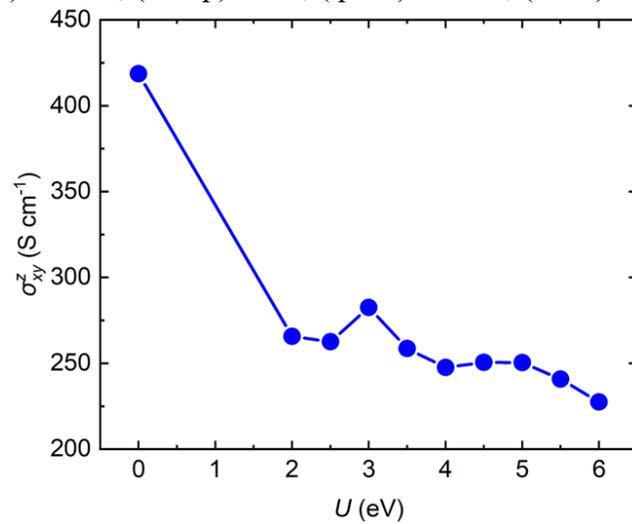

**Figure S17.** Theoretically calculated AHC for different values of Coulomb interaction (U).

# Reference

[1]   G. Kresse and J. Hafner, *Ab initio* molecular dynamics for liquid metals, Phys. Rev. B **47**, 558 (1993).

[2]   G. Kresse and J. Furthmüller, Efficient iterative schemes for *ab initio* total-energy calculations using a plane-wave basis set, Phys. Rev. B **54**, 11169 (1996).

[3]   J. P. Perdew, K. Burke, and M. Ernzerhof, Generalized Gradient Approximation Made Simple, Phys. Rev. Lett. **77**, 3865 (1996).

[4]   J. R. Yates, X. Wang, D. Vanderbilt, and I. Souza, Spectral and Fermi surface properties from Wannier interpolation, Phys. Rev. B **75**, 195121 (2007).

[5]   N. Nagaosa, J. Sinova, S. Onoda, A. H. MacDonald, and N. P. Ong, Anomalous Hall effect, Rev. Mod. Phys. **82**, 1539 (2010).

[6]   E. J. Telford, A.H. Dismukes, K. Lee, M. Cheng, A. Wieteska, A.K. Bartholomew, Y.S.Chen, X. Xu, A.N. Pasupathy, X. Zhu, C.R. Dean, X. Roy Layered Antiferromagnetism Induces Large Negative Magnetoresistance in the van der Waals Semiconductor CrSBr, Adv. Mater. **32**, 2003240 (2020).

[7]   H. Yamamoto, Y. Motomura, T. Anno, and T. Shinjo, Magnetoresistance of non-coupled [NiFe/Cu/Co/Cu] multilayers, J. Magn. Magn. Mater. **126**, 437 (1993).

[8]   S. Albarakati, C. Tan, Z.J. Chen, J.G. Partridge, G. Zheng, L. Farrar, E.L.H. Mayes, M.R. Field, C. Lee, Y. Wang, Y. Xiong, M. Tian, F. Xiang, A.R. Hamilton et al., Antisymmetric magnetoresistance in van der Waals Fe3GeTe2 /graphite/Fe3GeTe2 trilayer heterostructures, Sci. Adv. **5**, aaw0409 (2019).

[9]   D. Hou, G. Su, Y. Tian, X. Jin, S. A. Yang, and Q. Niu, Multivariable Scaling for the Anomalous Hall Effect, Phys. Rev. Lett. **114**, 217203 (2015).

[10]  L. Ye, Y. Tian, X. Jin, and D. Xiao, Temperature dependence of the intrinsic anomalous Hall effect in nickel, Phys. Rev. B **85**, 220403 (2012).

[11]  H. Lv, X. C. Huang, K. H. L. Zhang, O. Bierwagen, and M. Ramsteiner, Underlying Mechanisms and Tunability of the Anomalous Hall Effect in NiCo2O4 Films with Robust Perpendicular Magnetic Anisotropy, Adv. Sci. **10**, 2302956 (2023).

[12]  J. Chen, X. Yang, F. Zhou, Y.-C. Lau, W. Feng, Y. Yao, Y. Li, Y. Jiang, and W. Wang, Colossal anomalous Hall effect in the layered antiferromagnetic EuAl2Si2 compound, Mater. Horiz. **11**, 4665 (2024).